# Η ημερήσια μετακίνηση με σκοπό την εργασία ως πολύπλοκο δίκτυο: η περίπτωση της Ελλάδας

**Δημήτριος Τσιώτας, Κωνσταντίνος Ραπτόπουλος**

Τμήμα Μηχανικών Χωροταξίας Πολεοδομίας και Περιφερειακής Ανάπτυξης,
Πανεπιστήμιο Θεσσαλίας,
Πεδίον Άρεως, Βόλος, 38 334,
Τηλ +30 24210 74446, fax: +302421074493
E-mails: tsiotas@uth.gr; kosraptopoulos@uth.gr

**Περίληψη**
Αυτό το άρθρο μελετά το ελληνικό διαπεριφερειακό δίκτυο των ημερήσιων μετακινήσεων με σκοπό την εργασία (Greek commuting network - GRN), χρησιμοποιώντας μέτρα και τεχνικές της ανάλυσης σύνθετων δικτύων και εμπειρικές μεθόδους. Η μελέτη αποσκοπεί στην ανίχνευση των δομικών χαρακτηριστικών του φαινομένου commuting και του τρόπου με τον οποίο το δίκτυο αυτό υπηρετεί και προάγει την περιφερειακή ανάπτυξη. Στην εμπειρική ανάλυση κατασκευάζεται ένα πολλαπλό υπόδειγμα γραμμικής παλινδρόμησης για τον αριθμό των commuters, το οποίο δομείται με βάση το σημασιολογικό πλαίσιο της έννοιας του δικτύου, ως μία προσπάθεια προώθησης του διεπιστημονικού διαλόγου. Η ανάλυση αναδεικνύει την επίδραση των χωρικών περιορισμών στο δίκτυο, παρέχει πληροφορία σχετικά με τα μεγαλύτερα έργα υποδομής που συντελέστηκαν στον τομέα των οδικών μεταφορών και επηρέασαν τη μεταφορική ικανότητα της χώρας, και σκιαγραφεί τη βαρυτική διάσταση του φαινομένου, μέσα από την ελκτική δράση των πολυπληθών πόλεων που διατηρούν ένα μεγάλο ποσό δραστηριότητας commuting εντός των αστικών τους ορίων, περιορίζοντας έτσι τις διαπεριφερειακές διαρροές εργατικού δυναμικού και συνεπώς αυξάνοντας την ενδογενή τους παραγωγικότητα. Συνολικά, το άρθρο αναδεικνύει την αποτελεσματικότητα της χρήσης της ανάλυσης των σύνθετων δικτύων στη μοντελοποίηση των χωρικών δικτύων και ειδικότερα των συστημάτων μεταφορών και επιδιώκει την προώθηση χρήσης του παραδείγματος των δικτύων στις χωρικές και περιφερειακές εφαρμογές.

**Λέξεις Κλειδιά:** χωρικά δίκτυα, ανάλυση σύνθετων δικτύων, διαπεριφερειακή μετακίνηση με σκοπό την εργασία.

**Abstract**
This article studies the Greek interregional commuting network (GRN) by using measures and methods of complex network analysis and empirical techniques. The study aims to detect structural characteristics of the commuting phenomenon, which are configured by the functionality of the land transport infrastructures, and to interpret how this network serves and promotes the regional development. In the empirical analysis, a multiple linear regression model for the number of commuters is constructed, which is based on the conceptual framework of the term "network", in effort to promote the interdisciplinary dialogue. The analysis highlights the effect of the spatial constraints on the network's structure, provides information on the major road transport infrastructure projects that constructed recently and influenced the country capacity, and outlines a gravity pattern describing the commuting phenomenon, which expresses that cities of high population attract large volumes of commuting activity within their boundaries, a fact that contributes to the reduction of their outgoing commuting and consequently to the increase of their inbound productivity. Overall, this paper highlights the effectiveness of complex network analysis in the modeling of spatial and particularly of transportation network and promotes the use of the network paradigm in the spatial and regional research.

**1. Εισαγωγή**

Η *ημερήσια μετακίνηση με σκοπό την εργασία* (*commuting*) αποτελεί ένα πολυδιάστατο φαινόμενο που αφορά την τακτική κινητικότητα των εργαζομένων για εργασία, η οποία διενεργείται σε θέσεις εκτός των αστικών ορίων του τόπου διαμονής τους (Πολύζος, 2011). Το θεωρητικό πλαίσιο του φαινομένου εμπεριέχει κοινωνικές, οικονομικές, γεωγραφικές και πολιτικές διαστάσεις, με αποτέλεσμα η μελέτη και η σε βάθος γνώση του να συνιστά μία ιδιαίτερα πολύπλοκη διαδικασία που είναι δυνατό να παρέχει χρήσιμες ιδέες προς την κατεύθυνση άσκησης αποτελεσματικότερης πολιτικής, ιδιαίτερα στους τομείς της εργασίας και των μεταφορών, αλλά και στην προαγωγή του αειφόρου σχεδιασμού των μεταφορών (Evans et al., 2002; Van Ommeren and Rietveld, 2005). Μέχρι σήμερα, έχει μελετηθεί από τους Περιφερειολόγους και τους Συγκοινωνιολόγους μία ευρεία γκάμα θεμάτων commuting, όπως είναι ενδεικτικά το μεταφορικό (χωρικό και χρονικό) κόστος (Van Ommeren and Fosgerau, 2009; Tsiotas and Polyzos, 2013a) η ψυχολογία της μετακίνησης (Koslowsky et al., 1995), η πιθανότητα τροχαίου ατυχήματος (Ozbay et al., 2007), διάφορα ζητήματα επιλογής του τρόπου (μέσου) μετακίνησης (transportation modes) και των εναλλακτικών δυνατών διαδρομών (Murphy, 2009; Liu and Nie, 2011), καθώς και θέματα που αφορούν τη σχέση του φαινομένου με μορφές της παραγωγικότητας (Van Ommeren and Rietveld, 2005).

Ωστόσο, η μακροσκοπική μελέτη του commuting δεν έχει τύχει της ανάλογης προσοχής, τόσο διεθνώς όσο στην περίπτωση της Ελλάδας (Tsiotas and Polyzos, 2013a; Polyzos et al., 2014). Ένας από τους σύγχρονους επιστημονικούς τομείς που καθίσταται ικανός στην παροχή μεθόδων μοντελοποίησης προς αυτήν την ολιστική κατεύθυνση είναι η αποκαλούμενη *ανάλυση σύνθετων δικτύων* (complex network analysis) (Brandes and Erlebach, 2005; Easley και Kleinberg, 2010; Barthelemy, 2011) ή *Επιστήμη των Δικτύων* (Brandes et al., 2013), όπως έχει μετονομαστεί πρόσφατα. Αυτή η προσέγγιση αναπαριστά τα συστήματα επικοινωνίας ως γράφους (Easley and Kleinberg, 2010; Borgatti and Halgin, 2011; Tsiotas and Polyzos, 2013a), δηλαδή ως διμερή σύνολα που αποτελούνται από μία συλλογή διασυνδεδεμένων μονάδων (τους κόμβους) και από τις μεταξύ τους συνδέσεις (τις ακμές). Σύμφωνα με αυτήν την οπτική, ένα σύστημα ημερήσιας μετακίνησης εργαζομένων (commuters) μπορεί να αναπαρασταθεί ως δίκτυο (γράφος), στο οποίο, σε διαπεριφερειακή κλίμακα, οι κόμβοι εκφράζουν τις περιοχές προέλευσης και προορισμού και οι ακμές πληροφορίες απόστασης και ροών.

Εντός του παραπάνω εννοιολογικού πλαισίου, το άρθρο αυτό περιγράφει τη διαπεριφερειακή μετακίνηση commuting στην Ελλάδα ως πολύπλοκο δίκτυο, η οποία αναπτύσσεται μεταξύ των πρωτευουσών των χερσαίων νομών της χώρας. Τα χαρακτηριστικά του φαινομένου εξετάζονται τόσο μεμονωμένα, ως προς την τοπολογία και τη λειτουργικότητα του δικτύου που κατασκευάζεται, όσο και σε σχέση με το ευρύτερο κοινωνικοοικονομικό τους περιβάλλον. Περαιτέρω, στην εμπειρική ανάλυση προτείνεται ένα πολλαπλό υπόδειγμα γραμμικής παλινδρόμησης που περιγράφει τη διαπεριφερειακή ημερήσια μετακίνηση με σκοπό την εργασία, βασισμένο στις σημασιολογικές συνιστώσες της έννοιας του δικτύου, όπως αυτές περιγράφηκαν από τους Berners-Lee et al. (2007) και Easley and Kleinberg (2010) και αναθεωρήθηκαν από τους Tsiotas and Polyzos (2015c). Απώτερο σκοπό της μελέτης αποτελεί η ανίχνευση των δομικών χαρακτηριστικών του φαινομένου commuting, όπως αυτά αποτυπώνονται στις δυνατότητες μετακίνησης που παρέχονται μέσω των χερσαίων μεταφορικών υποδομών.

Το υπόλοιπο του εγγράφου δομείται ως εξής: στην 2η ενότητα παρουσιάζεται το μεθοδολογικό πλαίσιο και ειδικότερα ο τρόπος και οι παραδοχές μοντελοποίησης του διαπεριφερειακού δικτύου σε γράφο, τα μέτρα ανάλυσης δικτύου που χρησιμοποιούνται και το εμπειρικό υπόδειγμα που κατασκευάζεται. Στην 3η ενότητα παρουσιάζονται τα αποτελέσματα των επιμέρους αναλύσεων και σχολιάζονται υπό το πρίσμα της ανάλυσης σύνθετων δικτύων και της περιφερειακής επιστήμης, με έμφαση στον τομέα των μεταφορών. Τέλος, στην παράγραφο 4 αναπτύσσονται τα συμπεράσματα της μελέτης.

## 2. Μεθοδολογικό πλαίσιο
### 2.1. Μοντελοποίηση του Δικτύου

Το GCN (σχήμα 1) αποτελεί ένα δίκτυο με περισσότερο οικονομική και λιγότερο φυσική ερμηνεία. Αυτό το χωρικό μοντέλο αντιπροσωπεύει μία πτυχή του εθνικού οδικού δικτύου, εκφρασμένη σε διαπεριφερειακή κλίμακα (σε επίπεδο νομών), στην οποία δεν διατηρείται η πληροφορία της γεωμετρίας των οδών, αλλά μόνο η γεωγραφική κλίμακα των θέσεων των κόμβων. Με την κατασκευή του GCN ουσιαστικά επιχειρείται η αναπαράσταση των λειτουργιών και των σχέσεων οδικής επικοινωνίας που αναπτύσσονται μεταξύ των νομών της Ελλάδας, με σκοπό τη μελέτη της τοπολογίας και των οικονομικών δυναμικών που διαμορφώνονται από αυτό το σύστημα των χωρικών και οικονομικών αλληλεπιδράσεων.

Αναλυτικότερα, το GCN αναπαρίσταται στον *L*-χώρο (Barthelemy, 2011; Tsiotas and Polyzos, 2015a,b) ως ένας μη κατευθυνόμενος γράφος $G(V,E)$ με χωρικά βάρη (spatial network), του οποίου το σύνολο των κόμβων $V$ αντιπροσωπεύει τις *πρωτεύουσες των ελληνικών νομών*, ενώ το σύνολο των ακμών $E$ εκφράζει *την ύπαρξη δυνατότητας απευθείας οδικών συνδέσεων* μεταξύ των νομών της Ελλάδας. Οι θέσεις των κόμβων του GCN στο χάρτη (σχήμα 1) αντιστοιχούν στις γεωγραφικές συντεταγμένες των πρωτευουσών των ελληνικών νομών και τα μήκη των ακμών αναπαριστούν, υπό κλίμακα, τις ευκλείδειες χιλιομετρικές αποστάσεις των κόμβων. Η επιλογή του συγκεκριμένου είδους των κόμβων πραγματοποιείται λόγω της οικονομικής σημασίας που έχουν στην Περιφερειακή Επιστήμη οι πρωτεύουσες των νομών, ως χώροι σημαντικών πληθυσμιακών συγκεντρώσεων (Polyzos et al., 2013; Tsiotas and Polyzos, 2013a,b), με απώτερο σκοπό το χωρικό δίκτυο που θα προκύψει να αποτελεί ένα υπόδειγμα με σημαντικό οικονομικό αντίκρισμα.

Το GCN είναι *συνδετικό* (*connective*) δίκτυο (μία συνιστώσα), συναποτελούμενο από τους $n$=39 νομούς (κόμβους) της ηπειρωτικής χώρας και από τις $m$=71 μεταξύ τους οδικές συνδέσεις (ακμές) (σχήμα 1). Τα χωρικά βάρη $w_{s,ij}=d(e_{ij})$ των ακμών εκφράζουν τις πραγματικές χιλιομετρικές αποστάσεις των συντομότερων διαδρομών (σε χιλιόμετρα) που συνδέουν τις πρωτεύουσες των νομών. Κάθε ακμή αντιστοιχεί σε τμήματα διπλής οδικής κατεύθυνσης, με αποτέλεσμα ο πίνακας των συνδέσεων του δικτύου να προκύπτει συμμετρικός. Ως περαιτέρω βάρη στο GCN θεωρούνται οι χρονοαποστάσεις (spacetime distances) μεταξύ των κόμβων, οι οποίες εκφράζουν τον απαιτούμενο χρόνο (σε min) κάλυψης μιας δεδομένης χιλιομετρικής απόστασης μεταξύ δύο θέσεων του δικτύου. Οι τιμές αυτές αποτελούν και έναν έμμεσο δείκτη της αποτελεσματικότητας του οδικού διαπεριφερειακού δικτύου, καθόσον ο μέσος χρόνος διέλευσης μίας διαδρομής αντιπροσωπεύει την ποιότητα των οδικών υποδομών του δικτύου (Τσιώτας κά., 2012; Tsiotas and Polyzos, 2013a,b).

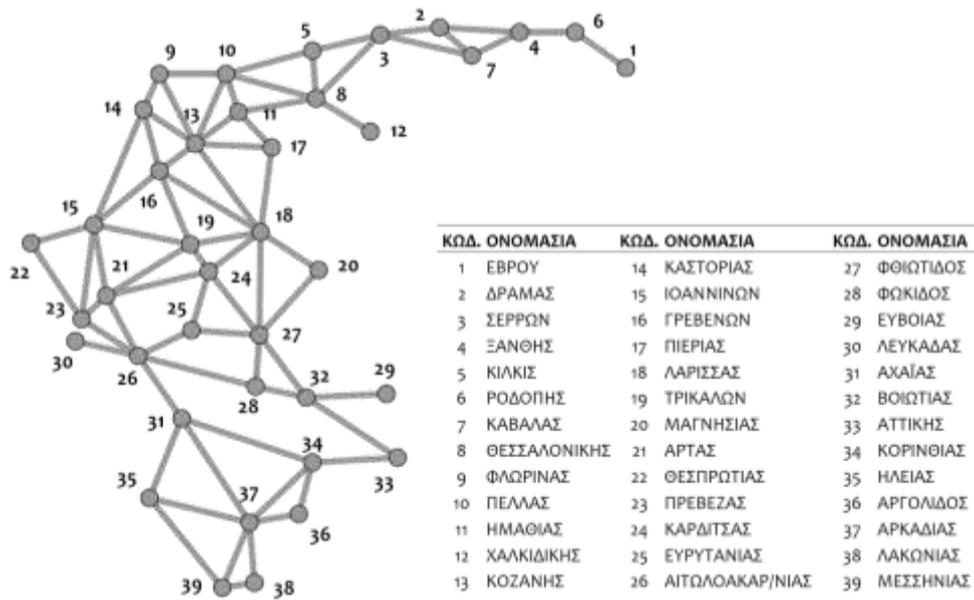

**Σχήμα 1.** Το διαπεριφερειακό δίκτυο των ημερησίως μετακινουμένων με σκοπό την εργασία (Greek Commuters Network - GCN), αναπαριστώμενο στον *L*-χώρο ως μη κατευθυνόμενος γράφος με *n*=39 κόμβους και *m*=71 ακμές (οι κόμβοι στο γράφο αντιπροσωπεύουν τις πρωτεύουσες των Καποδιστριακών νομών).

Τα χωρικά δεδομένα (γεωγραφικές συντεταγμένες) που χρησιμοποιούνται για την κατασκευή του GCN αντλήθηκαν από τις υπηρεσίες ψηφιακής χαρτογράφησης της Google (2013), ενώ τα δεδομένα των χιλιομετρικών αποστάσεων και των χρονοαποστάσεων από τις εργασίες των Τσιώτας κά. (2012) και Tsiotas and Polyzos (2013a,b) και (2015c). Τα διαθέσιμα δεδομένα των χρονοαποστάσεων αντιστοιχούν σε δύο χρονικές καταστάσεις (στιγμιότυπα) του διανομαρχιακού ελληνικού οδικού δικτύου. Η πρώτη περιλαμβάνει στοιχεία του έτους 1988, τα οποία περιγράφουν την κατάσταση του εθνικού δικτύου οδικών μεταφορών στο πιο πρόσφατο παρελθόν του, δηλαδή στο αρχικό στάδιο της σύγχρονης μορφής του. Η δεύτερη περιλαμβάνει στοιχεία του έτους 2010, τα οποία αντιπροσωπεύουν τη σημερινή εικόνα του δικτύου, έπειτα από την ενσωμάτωση στις οδικές υποδομές της χώρας ενός αριθμού έργων αναβάθμισης, όπως είναι ενδεικτικά η γέφυρα *Ρίου-Αντιρρίου* και η *Εγνατία Οδός*.

Τέλος, η πληροφορία που αφορά τους commuters του GCN ενσωματώθηκε στο χωρικό υπόδειγμα με τη μορφή βάρους κόμβων (node weight) και όχι βάρους ακμών (edge weight) (Tsiotas and Polyzos, 2013a,b). Δηλαδή, τα διαθέσιμα στοιχεία commuting δεν περιγράφουν τον αριθμό των μετακινούμενων με τη μορφή κυκλοφοριακού φόρτου (ροές commuters) στις ακμές του δικτύου, αλλά ως το πλήθος των εργαζομένων που ξεκινούν το ημερήσιο ταξίδι τους από μία συγκεκριμένη πρωτεύουσα νομού, με αποτέλεσμα το δίκτυο GCN να αποτελεί μη κατευθυνόμενο γράφο. Η περιγραφή της πληροφορίας commuting ως βάρη κόμβων θεωρήθηκε αντιπροσωπευτικότερη σε σχέση με την προσάρτηση βαρών στις ακμές του GCN, καθώς η πρώτη περιγράφει το σύνολο των μετακινούμενων με σκοπό την εργασία, ενώ η δεύτερη μόνο το ποσοστό των commuters που μετακινείται μεταξύ των πρωτευουσών των νομών.

*2.2. Μέτρα ανάλυσης δικτύου*

Τα μέτρα χώρου και τοπολογίας που χρησιμοποιούνται στην ανάλυση του GRN παρουσιάζονται συνοπτικά στον πίνακα 1 (βλ. Παράρτημα). Εκτός από τα βασικά αυτά μέτρα, στην ανάλυση του GCN υπολογίζεται ο *ωμέγα (ω) δείκτης* των Telesford et al.

(2011), με σκοπό την ανίχνευση της ιδιότητας του μικρού-κόσμου (small-world) *S-W* (Watts and Strogatz, 1998) και της ύπαρξης χαρακτηριστικών δικτυώματος (lattice-like characteristics) ή τυχαίου γράφου (random-like characteristics). Το μέτρο συγκρίνει τη μέση συγκέντρωση του εξεταζόμενου δικτύου $\langle c \rangle$ με αυτήν ενός ισοδύναμου δικτυώματος $\langle c \rangle_{latt}$ και το μέσο μήκος μονοπατιού $\langle l \rangle$ του δικτύου με το αντίστοιχο μέγεθος ενός ισοδύναμου τυχαίου γράφου $\langle l \rangle_{rand}$, με βάση τη σχέση:

$$\omega = \left(\frac{\langle l \rangle_{rand}}{\langle l \rangle}\right) - \left(\frac{\langle c \rangle}{\langle c \rangle_{latt}}\right) \tag{1}$$

Τιμές του ω δείκτη που βρίσκονται κοντά στο μηδέν περιγράφουν την ιδιότητα του μικρού-κόσμου, ενώ οι θετικές τιμές υποδηλώνουν την ύπαρξη τυχαίων χαρακτηριστικών στο δίκτυο και οι αρνητικές την ύπαρξη χαρακτηριστικών δικτυώματος (Tsiotas and Polyzos, 2015β). Τα μηδενικά πρότυπα (null models) που χρησιμοποιούνται για τον υπολογισμό της παραπάνω σχέσης δημιουργούνται με χρήση των αλγορίθμων παραγωγής τυχαίων γράφων, των Maslov and Sneppen (2002), και δικτυώματος, των Sporns and Kotter (2004), οι οποίοι είναι *επαναληπτικοί* (*iterative*) και διατηρούν την κατανομή βαθμού του πρότυπου (εμπειρικού) δικτύου. Ο πρώτος εφαρμόζεται σε δύο βήματα, αρχικά επιλέγονται τυχαία τέσσερις κόμβοι των οποίων οι ακμές διχοτομούνται, αντιστοιχίζοντας μισή ακμή σε κάθε κόμβο, και στη συνέχεια οι μισές ακμές ενώνονται με τυχαίο τρόπο μεταξύ τους (Rubinov and Sporns, 2010). Ο αλγόριθμος παραγωγής του ισοδύναμου δικτυώματος των Sporns and Kotter (2004) (*latticization algorithm*) εφαρμόζει την ίδια διαδικασία, θέτοντας τον περιορισμό ότι η εναλλαγή των μισών ακμών πραγματοποιείται μόνο όταν ο προκύπτων πίνακας συνδέσεων έχει τις μη μηδενικές του καταχωρήσεις εγγύτερα στην κύρια διαγώνιο σε σύγκριση με την αρχική του κατάσταση (Sporns and Kotter, 2004; Rubinov and Sporns, 2010). Με τη συνθήκη αυτή προσεγγίζεται η τοπολογία δικτυώματος, καθόσον στα δικτυώματα είναι απίθανο να πραγματοποιηθούν συνδέσεις απομακρυσμένων κορυφών (Sporns and Kotter (2004; Rubinov and Sporns, 2010).

Γενικά, η *S-W* ιδιότητα εξετάζεται με μαθηματική αυστηρότητα σε μια διαθέσιμη οικογένεια γράφων, όταν ανιχνευθεί πως το $\langle l \rangle$ δεν αυξάνεται γρηγορότερα από λογαριθμικά καθώς ο αριθμός των κόμβων τείνει στο άπειρο, όταν δηλαδή $\langle l \rangle_{bin} = \mathcal{O}(\log n)$ καθώς $n \rightarrow \infty$ (Porter, 2012). Επειδή δεν καθίσταται εφικτή η συλλογή μιας οικογένειας διαφορετικών διαχρονικών εκδοχών του GCN για τον έλεγχο της *S-W* ιδιότητας με την εφαρμογή του ορισμού (Tsiotas and Polyzos, 2015b), επιλέγεται η εξέταση της ιδιότητας του μικρού-κόσμου προσεγγιστικά, με χρήση του ω δείκτη. Η προσέγγιση αυτή παρέχει περαιτέρω ενδείξεις για το αν η τυπολογία του εξεταζόμενου δικτύου διέπεται από χαρακτηριστικά τυχαίου δικτύου (random network) ή δικτυώματος (lattice network).

*2.3. Εμπειρική ανάλυση*
Στην ενότητα αυτή κατασκευάζεται ένα εμπειρικό υπόδειγμα για τον υπολογισμό του αριθμού των ημερησίως μετακινουμένων του GCN, χρησιμοποιώντας μεταβλητές των κόμβων του δικτύου. Κάθε μεταβλητή αποτελεί, δηλαδή, συλλογή τιμών που εμφανίζουν οι κόμβοι του δικτύου σε ένα συγκεκριμένο χαρακτηριστικό *p*, η οποίες αποτελούνται από *n* το πλήθος στοιχεία (ίσα με τον αριθμό των κόμβων του δικτύου). Για παράδειγμα, το σύνολο **k**={$k_i$, $i=1,\ldots,n$} με τις τιμές που έχουν οι *n*=39 κόμβοι του GCN στο μέγεθος του βαθμού *k*, αντιμετωπίζεται ως μία *διανυσματική* στατιστική *μεταβλητή* (*vector variable*) βαθμού που ονομάζεται DEG (Tsiotas and Polyzos, 2013a, 2015c). Με δεδομένο ότι οι κόμβοι στο GCN αντιστοιχούν στους ελληνικούς νομούς, δημιουργούνται *p*=30 διανυσματικές μεταβλητές (*Y*, *X₁*,...,*X₂₉*) (όπου δεν υφίσταται κίνδυνος σύγχυσης θα καλούνται εφεξής απλώς *μεταβλητές*), οι οποίες συμμετέχουν στην εμπειρική ανάλυση. Η επιλογή του είδους των μεταβλητών πραγματοποιείται με κριτήριο την κατά το δυνατόν εξάντληση της θεματολογίας σχετικά με τη μελέτη του φαινομένου του commuting (Glaeser and Kohlhase, 2003; Clark et al., 2003; Ozbay et al., 2007; Van Ommeren and Fosgerau, 2009; Murphy, 2009; Liu and Nie, 2011; Polyzos, 2011; Tsiotas and Polyzos, 2015c), και ανάλογα με τη διαθεσιμότητα των στοιχείων. Περαιτέρω, οι μεταβλητές που χρησιμοποιούνται στην κατασκευή του υποδείγματος ομαδοποιούνται σε τρεις θεματικές κατηγορίες, σε *δομικές* (*structural*) *S*, *λειτουργικές* (*functional* or *behavioral*) *B* και *οντολογικές* (*ontological*) *O* μεταβλητές, ώστε να καλύπτουν και τις τρεις πτυχές που περιγράφουν εννοιολογικά ένα δίκτυο, όπως προτάθηκε από τους Tsiotas and Polyzos (2015c).

Με βάση τα παραπάνω, το σύνολο των 30 μεταβλητών που συμμετέχουν στην εμπειρική ανάλυση του GCN παρουσιάζονται στον πίνακα 2 (βλ. Παράρτημα). Στη συνέχεια κατασκευάζεται ένα εμπειρικό υπόδειγμα για τον προσδιορισμό του αριθμού των commuters που μετακινούνται εντός του GCN. Για την κατασκευή του υποδείγματος χρησιμοποιούνται ο *διμεταβλητός συντελεστής συσχέτισης του Pearson* (*Pearson's bivariate coefficient of correlation*) και η μέθοδος της *πολυμεταβλητής γραμμικής παλινδρόμησης* (*multivariate linear regression analysis*) (Norusis, 2004; Devore and Berk, 2012; Tsiotas and Polyzos, 2015c).

Ο αλγόριθμος της κατασκευής του υποδείγματος αποτελείται από τρία βήματα. Το πρώτο περιλαμβάνει την ομαδοποίηση των διαθέσιμων μεταβλητών, ανάλογα με τη θεματική τους συνάφεια, στις τρεις κατηγορίες (δομικές, συμπεριφορικές ή λειτουργικές και οντολογικές) του πίνακα 2. Η διαδικασία αυτή οδηγεί στη διαμόρφωση τριών συνόλων μεταβλητών ($\mathbf{X}_S$, $\mathbf{X}_B$ και $\mathbf{X}_O$), σύμφωνα με τη σχέση (Tsiotas and Polyzos, 2015b):

$$\mathbf{X} \equiv \{X_k, k=1,...,p\}$$
$$(\mathbf{X} \equiv \mathbf{X}_S \cup \mathbf{X}_B \cup \mathbf{X}_O) \wedge (\mathbf{X}_i \cap \mathbf{X}_j = \varnothing) \wedge (j \neq i, j = \{S,B,O\}) \quad (2)$$

όπου τα σύνολα $\mathbf{X}_S$, $\mathbf{X}_B$ και $\mathbf{X}_O$ αντιπροσωπεύουν τη *δομική*, τη *λειτουργική* και την *οντολογική* ομάδα αντίστοιχα.

Στο δεύτερο βήμα, ο αλγόριθμος ξεχωρίζει τις αντιπροσωπευτικότερες μεταβλητές ανά κατηγορία, χρησιμοποιώντας το διμεταβλητό συντελεστή συσχέτισης του *Pearson*, ο οποίος δίδεται από τη σχέση:

$$r(x,y) \equiv r_{xy} = \frac{\text{cov}(x,y)}{\sqrt{\text{var}(x) \cdot \text{var}(y)}} \equiv \frac{s_{xy}}{s_x \cdot s_y} \quad (3)$$

όπου το cov(*x*,*y*)≡$s_{xy}$ εκφράζει τη συμμεταβλητότητα των μεταβλητών *x*,*y*, ενώ τα $\sqrt{\text{var}(x)} \equiv s_x$, $\sqrt{\text{var}(y)} \equiv s_y$ αντιπροσωπεύουν τις τυπικές τους αποκλίσεις.

Ως αντιπροσωπευτικές μεταβλητές για την κάθε κατηγορία επιλέγονται αυτές που έχουν το μεγαλύτερο άθροισμα τετραγώνων των συντελεστών συσχέτισης, οι οποίοι υπολογίζονται στις στατιστικά σημαντικές (επιλεγμένο επίπεδο σημαντικότητας α≤10%) μεταβλητές, διακρίνοντας δύο περιπτώσεις: αρχικά υπολογίζονται αποκλειστικά μεταξύ των μεταβλητών που βρίσκονται εντός μίας συγκεκριμένης ομάδας $\mathbf{X}_{k=\{S,B,O\}}$ (within-groups calculations) και έπειτα για το *σύνολο των p=30 διαθέσιμων μεταβλητών* του πίνακα 10 (global calculations). Η μαθηματική έκφραση της διαδικασίας υπολογισμού των αντιπροσώπων κάθε ομάδας δίδεται από τη σχέση (Tsiotas and Polyzos, 2015c):

$$X_k \equiv \alpha\nu\tau\iota\pi\rho\delta\sigma\omega\pi\sigma\varsigma\,\{\mathbf{X}_k\}_{k=S,B,O} \equiv rep\{\mathbf{X}_k\}_{k=S,B,O}:$$
$$(X_k \in \mathbf{X}_k) \wedge (\forall X_i, X_j \in \mathbf{X}_k): \qquad(4)$$
$$\sum_i {}_k r^2(X_k, X_i) = \max\{\sum_i r^2(X_i, X_j) : P[r(X_i, X_j)=0] \leq 0{,}10\}$$

Στο τελευταίο στάδιο, οι μεταβλητές που επιλέγονται από την παραπάνω διαδικασία, ως αντιπρόσωποι των κατηγοριών $\mathbf{X}_S$, $\mathbf{X}_B$ και $\mathbf{X}_O$, τοποθετούνται ως ανεξάρτητες μεταβλητές ($X_i$) σε ένα υπόδειγμα πολυμεταβλητής γραμμικής παλινδρόμησης (Tsiotas and Polyzos, 2015c), με εξαρτημένη μεταβλητή $Y$ τον αριθμό των ημερησίως μετακινουμένων με σκοπό την εργασία (commuters) (πίνακας 2).

Με το υπόδειγμα της γραμμικής παλινδρόμησης εκτιμάται η μορφή της γραμμικής σχέσης που περιγράφει καλύτερα τη σχέση μεταξύ της εξαρτημένης και των ανεξάρτητων μεταβλητών, η οποία στηρίζεται στη μέθοδο βελτιστοποίησης *των ελαχίστων τετραγώνων* (*ordinary least squares method*), υπό τον περιορισμό ότι τα τυποποιημένα λάθη ακολουθούν την κανονική κατανομή (Norusis, 2004). Σύμφωνα με τα προαναφερόμενα, το εμπειρικό υπόδειγμα που κατασκευάστηκε για τον αριθμό των μετακινουμένων στο διαπεριφερειακό δίκτυο commuting της Ελλάδας περιγράφεται από την παρακάτω σχέση (Tsiotas and Polyzos, 2015c):

$$\mathbf{X} \equiv \{X_k, k=1,...,p\}$$
$$(\mathbf{X} \equiv \mathbf{X}_S \cup \mathbf{X}_B \cup \mathbf{X}_O) \wedge (\mathbf{X}_i \cap \mathbf{X}_j = \varnothing) \wedge (j \neq i, j = \{S,B,O\})$$
$$GCN(V,E):$$
$$Y = f(X_{structural}, X_{functional}, X_{ontological}) =$$
$$= f(X_S, X_B, X_O) = b_S \cdot X_S + b_B \cdot X_B + b_O \cdot X_O + c \qquad(5)$$
$$\text{όπου } X_k \equiv rep\{\mathbf{X}_k\}_{k=S,B,O}:$$
$$X_k \in \mathbf{X}_k \wedge \forall X_i, X_j \in \mathbf{X}_k:$$
$$\sum_i {}_k r^2(X_k, X_i) = \max\{\sum_i r^2(X_i, X_j) : P[r(X_i, X_j)=0] \leq 0{,}10\}$$

### 3. Ανάλυση της τοπολογίας του δικτύου
*3.1. Υπολογισμός των μέτρων δικτύου (network measures)*
Στο πρώτο στάδιο της ανάλυσης υπολογίζονται τα μέτρα δικτύου του GCN, τα αποτελέσματα των οποίων παρουσιάζονται συγκεντρωτικά στον πίνακα 3.

**Πίνακας 3**
Συγκριτικός πίνακας με τα αποτελέσματα του υπολογισμού των μέτρων δικτύου για το GCN και το GRN

| Μετρική/ Μέγεθος | Σύμβολο | Μονάδα | Τιμή GCN |
|---|---|---|---|
| Αριθμός κόμβων | $n$ | #[a] | 39 |
| Αριθμός ακμών | $m$ | # | 71 |
| Κόμβοι με αυτοσυνδέσεις | $n(e_{ii} \in E)$ | # | 0 |
| Πλήθος απομονωμένων κόμβων | $n_{k=0}$ | # | 0 |
| Συνδετικές συνιστώσες | $\alpha$ | # | 1 |
| Μέγιστος βαθμός κόμβων | $k_{max}$ | # | 7 |
| Ελάχιστος βαθμός κόμβων | $k_{min}$ | # | 1 |
| Μέσος βαθμός κόμβων | $\langle k \rangle$ | # | 3,641 |

| Μετρική/ Μέγεθος | Σύμβολο | Μονάδα | Τιμή GCN |
|---|---|---|---|
| Μέσος (χωρικά) σταθμισμένος βαθμός κόμβων | $\langle k_w \rangle$ | km | 322,264 |
| Μέσος βαθμός εγγύτερων γειτόνων | $\langle k_{N(v)} \rangle$ | # | 3,641 |
| Μέσος σταθμισμένος βαθμός εγγύτερων γειτόνων | $\langle k_{N(v),w} \rangle$ | km | 322,26 |
| Μέσο μήκος ακμών | $\langle d(e_{ij}) \rangle$ | km | 85,497 |
| Συνολικό μήκος ακμών | $\sum_{ij} d(e_{ij})$ | km | 3.334,4 |
| Μέσο μήκος μονοπατιού | $\langle l \rangle$ | # | 4,58 |
| Μέσο μήκος μονοπατιού | $d(\langle l \rangle)$ | km | 389,045 |
| Διάμετρος δικτύου (δυαδική) | $d_{bin}(G)$ | # | 14 |
| Μήκος διαμέτρου δικτύου | $d_w(G)$ | km | 1.124,4 |
| Πυκνότητα γράφου (επίπεδου) | $\rho$ | net[d] | 0,640 |
| Πυκνότητα γράφου (μη επίπεδου) | $\rho$ | net | 0,097 |
| Συντελεστής συγκέντρωσης[c] | $C$ | net | 0,47 |
| Μέσος συντελεστής συγκέντρωσης[c] | $\langle C \rangle$ | net | 0,422 |
| Συναρμολογησιμότητα | $Q$ | net | 0,566 |

a. Πλήθος στοιχείων
b. NaN = not a number (απροσδιοριστία)
c. n/a = not available (μη διαθέσιμο)
d. Αδιάστατος αριθμός

Εξ' ορισμού, το GCN δε διαθέτει κόμβους με αυτοσυνδέσεις ($n(e_{ii} \in E)$=0), ούτε απομονωμένους κόμβους (isolated nodes - $n_{k=0}$), ούτε περισσότερες από μία συνιστώσες ($a_{GCN}$=1). Ο μέγιστος βαθμός του GCN είναι $k_{GCN,max}$=7, ενώ η ελάχιστη τιμή του $k_{GCN,min}$=1 και οφείλεται προφανώς στη συνεκτικότητά του (δηλαδή στην απουσία κόμβων με μηδενικό βαθμό). Περαιτέρω, η μέση τιμή του βαθμού του διαπεριφερειακού δικτύου ισούται με $\langle k \rangle_{GCN}$=3,641 και είναι κοντά στην περιοχή που εμφανίζεται η μεγαλύτερη συχνότητα τιμών της κατανομής βαθμού των αστικών οδικών συστημάτων, σύμφωνα με τη μελέτη των Courtat et al. (2010). Το μέσο μήκος μονοπατιού (average path length) εκφράζει γενικά το χωρικό κόστος (σε πλήθος ακμών) που απαιτείται για τη διενέργεια των μετακινήσεων στο εν λόγω δίκτυο (Tsiotas and Polyzos, 2015a,b). Για το GCN το κόστος αυτό υποδηλώνει ότι η διαδρομή που πρέπει να διανυθεί μεταξύ δύο τυχαίων κόμβων του δικτύου αποτελείται από $\langle l \rangle_{GCN}$=4,58 χωρικές μονάδες (ακμές ή βήματα διαχωρισμού).

Η τιμή του $\langle l \rangle_{GCN}$ βρίσκεται κοντά στη τάξη μεγέθους $\mathcal{O}(\sqrt{n}) = \sqrt{39} \approx 6{,}245$ του μέσου μήκους μονοπατιού ενός ισοκομβικού δικτυώματος $\langle l \rangle_{latt}$, παρέχοντας ενδείξεις για τη συνάφεια του GCN με το θεωρητικό πρότυπο. Επιπρόσθετα, η χωρική (χιλιομετρική) εκδοχή του μέσου μήκους μονοπατιού του GCN ισούται με $d(\langle l \rangle)_{GCN}$=389,045km και εκφράζει τη μέση χιλιομετρική απόσταση που απαιτείται για να διανυθούν τυχαία δύο κόμβοι στο δίκτυο. Ακολούθως, το μέγεθος της δυαδικής (τοπολογικής) διαμέτρου εκφράζει ότι η πιο απομακρυσμένη δυαδική οδική απόσταση που δύναται να διανυθεί διαπεριφερειακά αποτελείται από 14 ακμές, ενώ απέχει $d$(GCN)=1.124,40km (δίχως να υφίσταται απαραίτητα ταύτιση των δύο). Η τιμή της πυκνότητας $ρ$ του GCN, θεωρούμενου ως επίπεδου γράφου (αφαιρουμένων των ανισόπεδων κόμβων) ισούται με $ρ_{1,GCN}$=0,64, ενώ για τη μη επίπεδη περίπτωση (συμπεριλαμβανομένων των ανισόπεδων κόμβων) ισούται με $ρ_{2,GCN}$=0,097, τιμές που είναι εξαιρετικά μικρές σε σχέση με τις αντίστοιχες εμπειρικές τιμές των αστικών οδικών δικτύων (Barthelemy, 2011).

Η τιμή του συντελεστή συγκέντρωσης (clustering coefficient) του GCN (αριθμός τριγώνων/ αριθμός τριπλετών στο δίκτυο) ισούται με $C_{GCN}$=0,47 και υποδηλώνει ικανοποιητική ομαδοποίηση στη δομή του δικτύου. Περαιτέρω, ο μέσος συντελεστής συγκέντρωσης (ο αντίστοιχος λόγος υπολογισμένος στη γειτονιά κάθε κόμβου) ισούται με $\langle C \rangle_{GCN}$=0,422, ο οποίος προκύπτει εντυπωσιακά μεγαλύτερος από την αντίστοιχη τιμή ενός τυχαίου δικτύου ER $\langle C \rangle_{ER}$~ 1/$n$=1/39=0,026, εκφράζοντας ότι το δίκτυο απέχει από το να αποτελεί αποτέλεσμα τυχαίων διεργασιών.

Τέλος, η τιμή της συναρμολογησιμότητας (modularity) του GCN ισούται με $Q_{GCN}$=0,566, εκφράζοντας την ικανότητα επιμερισμού του δικτύου σε κοινότητες. Η τιμή αυτή περιγράφει μία ικανοποιητική ικανότητα διαίρεσης σε κοινότητες, καλύτερη τουλάχιστον από τις περιπτώσεις διαμερισμού οδικών δικτύων, οι οποίες στην πράξη εμφανίζονται συνήθως της τάξεως του $Q_{bipart}$<0,4.

*3.2. Μελέτη της τοπολογίας του διαπεριφερειακού δικτύου των commuters της Ελλάδας*
Για τη μελέτη της κατανομής του βαθμού (degree distribution) των κόμβων του GCN εξετάζονται από τα διαγράμματα διασποράς ($k$, $n(k)$) του σχήματος 2. Τα διαγράμματα αυτά εμφανίζουν οξυμμένο πρότυπο (peaked distribution), η τυπολογία του οποίου διαφοροποιείται από το πρότυπο κανόνα-δύναμης (power-law). Επίσης, η όξυνση (peak) που παρατηρείται γύρω από τη μέση τιμή $\langle k \rangle_{GCN}$~3 της κατανομής υποδηλώνει την παρουσία ισχυρών χωρικών περιορισμών (Barthelemy, 2011) στη δομή του GCN.

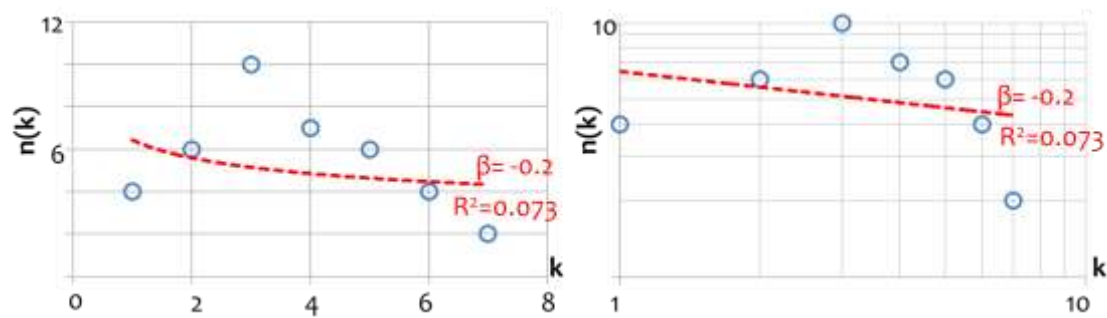

**Σχήμα 2.** Διαγράμματα διασποράς ($k$, $n(k)$) της κατανομής βαθμού του GCN σε μετρική (αρ.) και λογαριθμική (δεξ.) κλίμακα.

Στο επόμενο στάδιο, εξετάζονται τα διαγράμματα σποραδικότητας (spy plots) (σχήμα 3) (Χατζίκος, 2007) (a) του πίνακα συνδέσεων του GCN και τεσσάρων ισοκομβικών ($n$=39=σταθ) μηδενικών προτύπων (null models), με τις ιδιότητες (b) *ελευθέρου κλίμακας* (scale-free network), (c) *δικτυώματος* (lattice network), (d) *μικρού κόσμου* (small-world) και (e) *τυχαίου δικτύου* (random network) αντίστοιχα. Από τη σύγκριση των διαγραμμάτων, προκύπτει εμφανώς ότι η τυπολογία του προτύπου σποραδικότητας του GCN είναι παρόμοια με αυτή του (c) ισοκομβικού δικτυώματος (lattice network), αλλά οι τιμές του πίνακα συνδέσεων του GCN εμφανίζονται ελαφρώς πιο απομακρυσμένες από την κύρια διαγώνιο, σε σχέση με την πρότυπη περίπτωση.

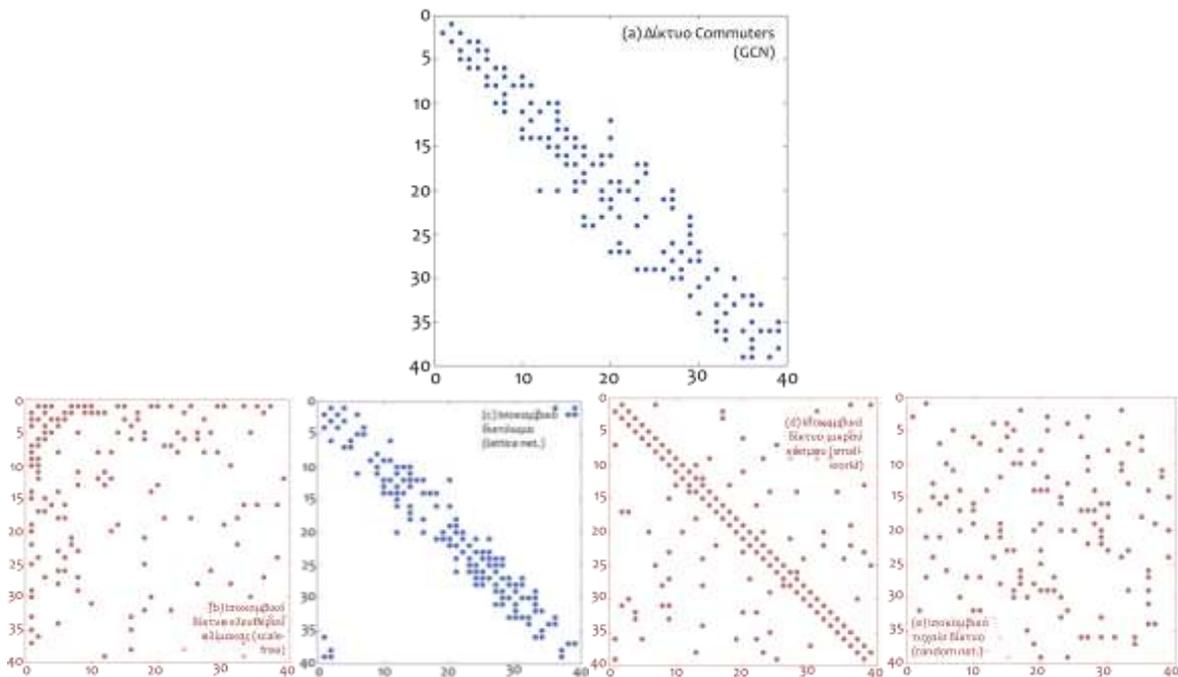

**Σχήμα 3.** Διαγράμματα σποραδικότητας (spy plots) των πινάκων συνδέσεων (adjacency matrices) (a) του οδικού δικτύου της Ελλάδας (GRN) (b) Ενός ισοκομβικού δικτύου με την ίδια κατανομή βαθμού και την ιδιότητα ελευθέρου κλίμακας (scale-free) (c) Ενός ισοκομβικού δικτυώματος (lattice network) με την ίδια κατανομή βαθμού (d) Ενός ισοκομβικού δικτύου με την ίδια κατανομή βαθμού και την ιδιότητα του μικρού κόσμου (small-world) και (e) Ενός ισοκομβικού τυχαίου δικτύου (random network) με την ίδια κατανομή βαθμού.

Η ανάλυση των διαγραμμάτων σποραδικότητας φαίνεται να επαληθεύεται από τα αποτελέσματα υπολογισμού του *ωμέγα* (*ω*) *δείκτη* (Telesford et al., 2011), τα οποία παρουσιάζονται στον πίνακα 4. Όπως προκύπτει, το GCN έχει χαρακτηριστικά δικτυώματος (lattice-like characteristics), γεγονός το οποίο είναι αναμενόμενο για περιπτώσεις δικτύων που υποβάλλονται σε έντονους χωρικούς περιορισμούς.

**Πίνακας 4**
Αποτελέσματα της προσεγγιστικής ανάλυσης ανίχνευσης της ιδιότητας μικρού-κόσμου (small-world) για το GCN

| Μέγεθος | $\langle c \rangle$ | $\langle c \rangle_{latt}$ | $\langle l \rangle$ | $\langle l \rangle_{rand}$ | $\omega^*$ |
|---|---|---|---|---|---|
| GCN | 0.422 | 0.312 | 4.580 | 2.889 | **-0.7218** |
| Ένδειξη | Συμπεριφορά characteristics) | | δικτυώματος | | (lattice-like |

*. Σύμφωνα με τη σχέση (1)

Στο επόμενο βήμα υπολογίζονται τα βασικά μέτρα τοπολογίας και κεντρικότητας (βαθμός, ενδιαμεσότητα, εγγύτητα, συγκέντρωση, συναρμολογησιμότητα και χωρική ισχύς) κόμβων του GCN, τα οποία παρουσιάζονται στις χωρικές κατανομές που απεικονίζονται στους τοπολογικούς χάρτες του σχήματος 4. Αρχικά, εξετάζεται η χωρική κατανομή του βαθμού (*k*) (σχήμα 4.a), η οποία σχηματίζει ένα ευδιάκριτο πρότυπο, με μία συστάδα ισχυρά συνδεδεμένων κόμβων που τοποθετείται στον κεντρικό κορμό του δικτύου commuters, αλλά και μία μεμονωμένη πλήμνη που βρίσκεται στο υποδίκτυο της Πελοποννήσου. Η συστάδα του κεντρικού κορμού διαμορφώνεται με βασικές τις πλήμνες των νομών Λάρισας, Λαμίας, Κοζάνης, Αιτωλοακαρνανίας και Ιωαννίνων, ενώ η πλήμνη της Πελοποννήσου βρίσκεται στο νομό Αρκαδίας.

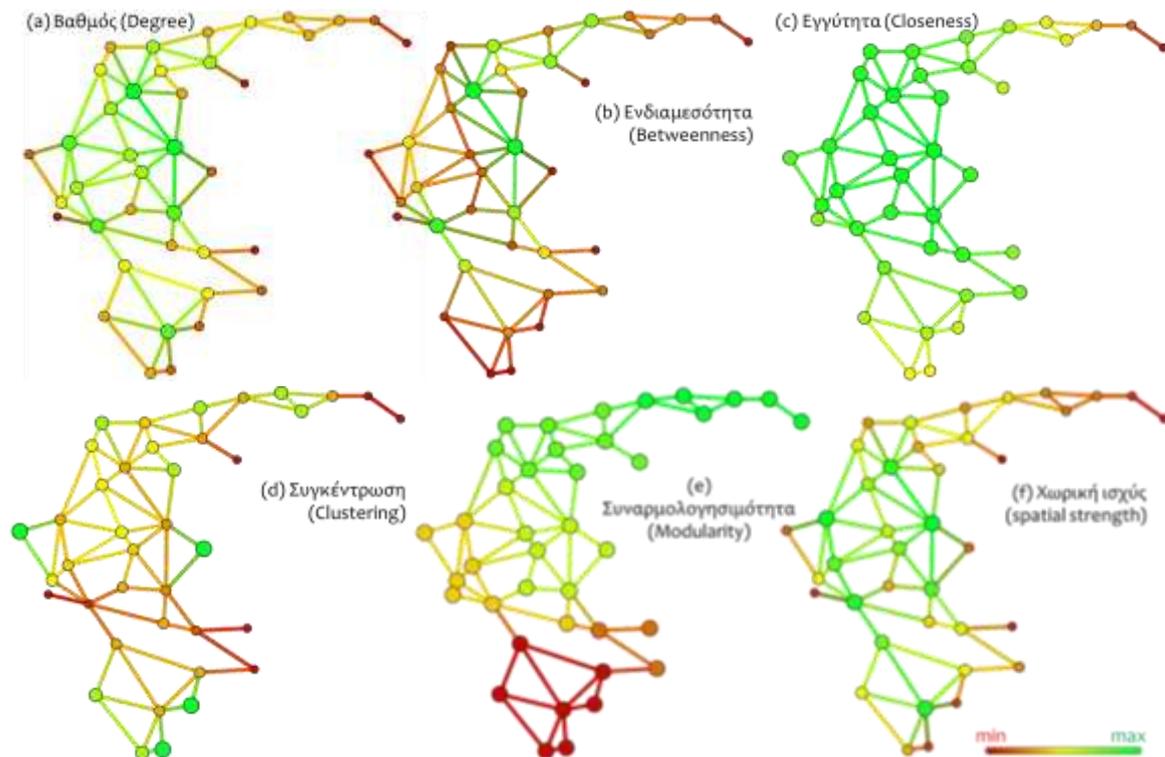

**Σχήμα 4.** Χωρική κατανομή των μέτρων κόμβου για το GCN: (a) Βαθμός (degree) (b) Ενδιαμεσότητα (betweenness) (c) Εγγύτητα (closeness) (d) Συγκέντρωση (clustering) (e) Συναρμολογησιμότητα (modularity classification) και (f) Χωρική ισχύς (spatial strength).

Δευτερευόντως, αξιοσημείωτη συνδετικότητα παρουσιάζουν οι νομοί της Πέλλας και της Θεσσαλονίκης στη Βόρεια Ελλάδα, καθώς και οι νομοί που διαμορφώνουν το τόξο Γρεβενά-Τρίκαλα-Καρδίτσα-Άρτα στην Κεντρική Ελλάδα. Λαμβάνοντας υπόψη ότι το μέγεθος του βαθμού εκφράζει τη συνδετικότητα και συνεπώς τη δυνατότητα επικοινωνίας των κόμβων του δικτύου, προκύπτει ότι η χωρική κατανομή του βαθμού (σχήμα 4.a) επισημαίνει τους κόμβους του GCN που εμφανίζουν πλεονέκτημα σύνδεσης έναντι των υπολοίπων. Το πλεονέκτημα αυτό οφείλεται στη γεωγραφική διευθέτηση των νομών της Ελλάδας, η οποία ευνοεί τη δημιουργία συνδέσεων περισσότερο στους κεντρικούς κόμβους και λιγότερο στους περιφερειακούς.

Έπειτα, η χωρική κατανομή της ενδιαμέσου κεντρικότητας (betweenness centrality) $C^b$ (σχήμα 4.b) παρουσιάζει μεγαλύτερη ένταση των μεγιστοβάθμιων τιμών στην ανατολική πλευρά της χώρας, στην οποία υφίστανται σαφέστερα περισσότερο αναβαθμισμένες υποδομές (Τσιώτας κά, 2012). Αντίθετα, η κατανομή των τιμών της κεντρικότητας εγγύτητας (closeness centrality) $C^c$ (σχήμα 4.c) παρουσιάζει μικρές τιμές στις μεθόριες περιοχές (Ανατολική Μακεδονία, Θράκη, Δυτική Πελοπόννησος), ενώ οι μεγάλες τιμές συγκεντρώνονται στον κεντρικό (ηπειρωτικό) κορμό της χώρας, επιβεβαιώνοντας το σαφές πλεονέκτημα που έχουν οι κεντρικές περιοχές στα χωρικά δίκτυα.

Ακολούθως, η χωρική κατανομή του συντελεστή συγκέντρωσης (clustering coefficient) $C$ (σχήμα 4.d) παρουσιάζει τους μεγιστοβάθμιους κόμβους να διατάσσονται στην περιφέρεια του διοικητικού πλέγματος, ήτοι στους νομούς *Ηλείας*, *Μεσσηνίας*, *Λακωνίας* και *Αργολίδας* στην Πελοπόννησο, στους νομούς *Θεσπρωτίας* και *Μαγνησίας* στο κεντρικό τμήμα της χώρας και στους νομούς *Πιερίας*, *Φλώρινας*, *Κιλκίς*, *Δράμας* και *Καβάλας* στο βόρειο τμήμα της Ελλάδας. Η κατάσταση αυτή εκφράζει γενικά ότι οι περιφερειακοί νομοί της χώρας έχουν μεγαλύτερη πιθανότητα να σχετίζονται με αλληλοσυνδεδεμένους γείτονες, περιγράφοντας το προνόμιο των πρώτων να απολαμβάνουν κατά τις αλληλεπιδράσεις τους στο δίκτυο πληροφορία μεγαλύτερης συνάφειας στο περιεχόμενό της. Όμως, αυτό το προνόμιο αρκετές φορές μετατρέπεται σε μειονέκτημα, διότι υποδηλώνει την εξάρτηση των εν λόγω κόμβων στους γείτονές τους, ως προς την ποικιλία των εισερχόμενων σημάτων. Για το GCN η ικανότητα πρόσβασης στο δίκτυο των κόμβων με μεγάλο συντελεστή συγκέντρωσης εξαρτάται από τις υποδομές οδικής επικοινωνίας των γειτόνων τους, οι οποίες, λόγω του υψηλού βαθμού αλληλεξάρτησης μεταξύ των γειτόνων, ενδέχεται να εμφανίζουν παρόμοια ποιοτικά χαρακτηριστικά.

Έπειτα, η χωρική κατανομή των τιμών της συναρμολογησιμότητας και ακριβέστερα της *Q-κατηγοριοποίησης* (modularity classification) (σχήμα 4.e) παρουσιάζεται συνεπής τόσο με τη θεωρία (Guimera et al., 2005; Kaluza et al., 2010; Barthelemy, 2011). Ειδικότερα, η κατανομή αυτή ακολουθεί έναν ευδιάκριτο επιμερισμό σε ζώνες γεωγραφικής συνάφειας, ο οποίος είναι αναμενόμενος για ένα δίκτυο με χαρακτηριστικά δικτυώματος, όπως είναι το διαπεριφερειακό των commuters.

Τέλος, η κατανομή της χωρικής ισχύος (spatial strength) $s$ (σχήμα 4.f) παρουσιάζεται εντατικότερη στο κέντρο, σχηματίζοντας μία διάταξη σε μορφή «πετάλου» που αποτελείται από τους νομούς *Φθιώτιδας*, *Λάρισας*, *Κοζάνης*, *Ιωαννίνων*, *Άρτας*, *Αιτωλοακαρνανίας* και *Αρκαδίας*. Αυτή η παρατήρηση οφείλεται προφανώς στο γεγονός ότι το μικρό πλήθος κόμβων και ακμών του GCN, σε συνδυασμό με τη διατήρηση της γεωγραφικής του κλίμακας, προσδίδει στα μήκη των ακμών την ίδια, περίπου, τάξη μεγέθους, με αποτέλεσμα η ισχύς $s=f(k,d(e))$ να εξαρτάται περισσότερο από το βαθμό $k$ και λιγότερο από τις αποστάσεις $d(e)$.

Έχοντας διαθέσιμα για το GCN εκτός από τα χωρικά δεδομένα και τα στοιχεία των χρονοαποστάσεων, μελετάται ακολούθως συγκριτικά (σχήμα 5) η διαχρονική μεταβολή της χωρικής κατανομής των μεγεθών (a) της συνδετικότητας (βαθμός $k$) και (b) της εγγύτητας (κεντρικότητα εγγύτητας $C^c$) του διαπεριφερειακού δικτύου commuters, για τα έτη 1988 και 2010.

Αρχικά, στο σχήμα 5(a) παρουσιάζεται η διαχρονική (1988 *vs* 2010) χωρική μεταβολή του βαθμού ($k$), η οποία εκφράζει τη μεταβολή στη συνδεσιμότητα του διαπεριφερειακού διοικητικού δικτύου της Ελλάδας κατά τις δύο αυτές περιόδους. Από τη σύγκριση των χρονικών στιγμιότυπων, η μοναδική διαφοροποίηση στη συνδετικότητα του δικτύου διακρίνεται για τους νομούς Αιτωλοακαρνανίας και Αχαΐας, οι οποίοι αύξησαν το βαθμό τους κατά μία σύνδεση. Η μεταβολή αυτή οφείλεται προφανώς στην κατασκευή της γέφυρας *Ρίου-Αντιρρίου*, η οποία προσέδωσε άμεση οδική πρόσβαση στους εν λόγω νομούς (Τσιώτας κά., 2012).

Στη συνέχεια, η χωρική κατανομή της κεντρικότητας εγγύτητας του έτους 1988 (σχήμα 5.a) περιγράφει την προσβασιμότητα του δικτύου μεταφορών εκείνης της περιόδου, η οποία εμφανίζεται γενικά δυσχερής για τους ακριτικούς νομούς. Οι περισσότερο προνομιούχοι σε συνδέσεις νομοί το 1988 φαίνεται πως ήταν οι *Κοζάνης*, *Λάρισας*, *Ιωαννίνων* και *Αρκαδίας*, προφανώς λόγω της κεντρικής γεωγραφικής τους θέσης, ενώ σε δευτερεύουσα θέση βρίσκονται οι νομοί της ευρύτερης *Κεντρικής Ελλάδας* και της *Θεσσαλονίκης*. Ιδιαίτερα, ο νομός της *Θεσσαλονίκης* φαίνεται πως κατείχε αξιοσημείωτο ρόλο στην τότε προσβασιμότητα, διότι αποτελούσε ήδη από το 1988 μητροπολιτικό πυρήνα της χώρας (Τσιώτας κά., 2012).

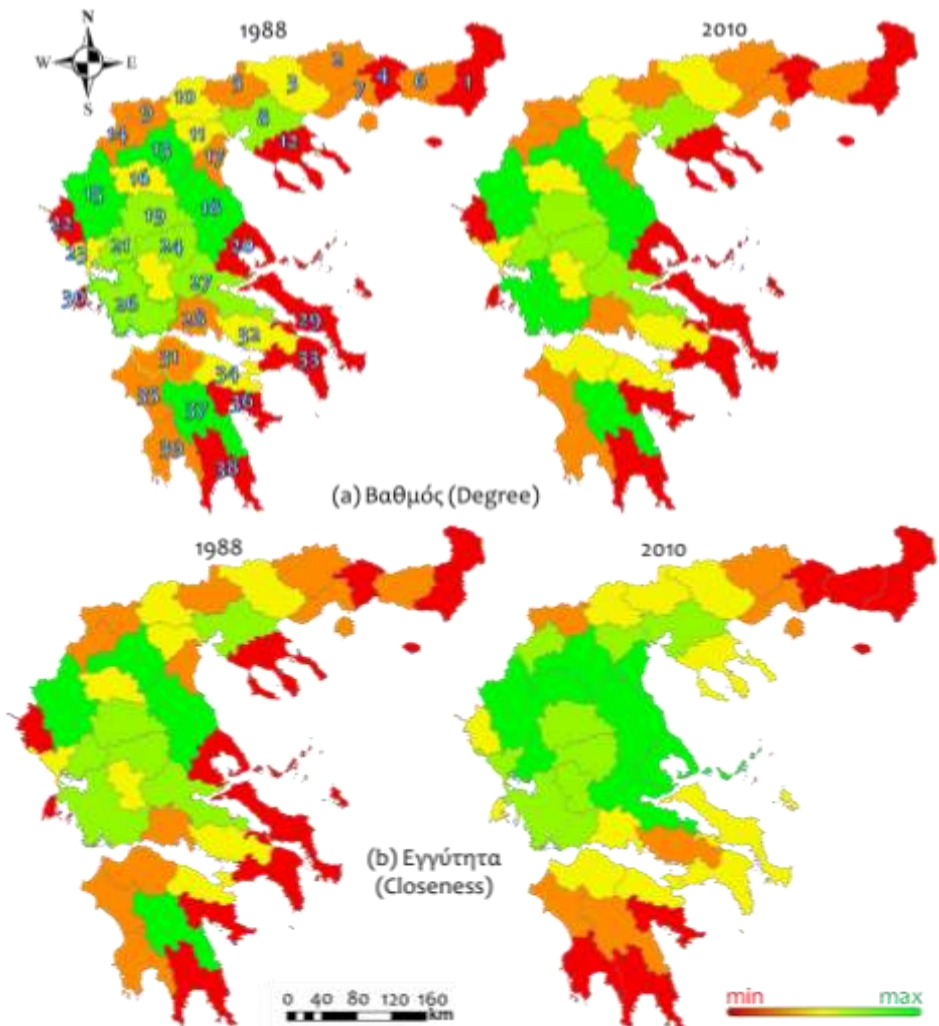

**Σχήμα 5.** Χωρική κατανομή των κεντρικοτήτων (a) βαθμού (degree centrality) και (b) εγγύτητας (closeness centrality) του ελληνικού διαπεριφερειακού δικτύου commuters (GCN), κατά τα έτη 1988 και 2010 (στην περίπτωση της εγγύτητας η χρωματική διαβάθμιση απεικονίζει σχετικές θέσεις και όχι απόλυτες τιμές κεντρικότητας) (πηγή: Τσιώτας κά., 2012).

Από τη σύγκριση των δύο χρονικών στιγμιότυπων (1988 *vs* 2010) της χωρικής κατανομής της κεντρικότητας εγγύτητας σκιαγραφείται η ποιοτική μεταβολή που υπέστησαν οι μεταφορικές υποδομές της Ελλάδας στο ενδιάμεσο χρονικό διάστημα (Τσιώτας κά., 2012). Όπως προκύπτει, η αναβάθμιση του οδικού δικτύου που συντελέστηκε την περίοδο 1988-2010 ευνόησε την ευρύτερη *Κεντρική Ελλάδα*, στην οποία συσπειρώνονται οι περιοχές με τη μεγαλύτερη προσβασιμότητα. Η ομοιομορφία που εμφανίζεται στο πρότυπο του 2010 υποδηλώνει τη σύγκλιση της μεταβλητότητας στα ποιοτικά χαρακτηριστικά των οδικών υποδομών της χώρας, γεγονός που εκφράζει ότι η προσβασιμότητα στη σημερινή μορφή του δικτύου αποτελεί κυρίως θέμα γεωγραφικής θέσης και λιγότερο υποδομών.

Ιδιαίτερο ενδιαφέρον παρουσιάζει η μείωση της σχετικής θέσης των νομών *Ροδόπης* και *Βοιωτίας*, καθεμία εκ των οποίων αποδίδεται σε διαφορετικούς λόγους. Αφενός, η μείωση που παρατηρείται στην περίπτωση της *Ροδόπης* οφείλεται προφανώς στην ακριτική γεωγραφική θέση του νομού, καθόσον η προσβασιμότητα του οδικού δικτύου της χώρας αυξήθηκε σε απόλυτα μεγέθη για το σύνολο των τμημάτων του. Αφετέρου, η μείωση της σχετικής θέσης στην προσβασιμότητα του νομού *Βοιωτίας* σχετίζεται πιθανότατα με την ανταγωνιστική δράση του νομού Αττικής, η οποία σκιαγραφεί μία σχέση βαρυτικής αλληλεξάρτησης μεταξύ αυτών των δύο νομών, με κυρίαρχο τον πολυπληθέστερο νομό Αττικής.

Περαιτέρω, στο σχήμα 6 παρουσιάζεται η χωρική κατανομή (a) της κεντρικότητας ευθύτητας ($C^s$) για το έτος 2010 και (b) της διαφοράς των μέσων χρονοαποστάσεων ανά νομό για τις περιόδους 1988 και 2010. Αρχικά, η χωρική κατανομή των τιμών της κεντρικότητας ευθύτητας (σχήμα 6.a) εκφράζει το ποσοστό της απόκλισης από την πλήρη ευθυγραμμία που εμφανίζει η οδική πρόσβαση ενός κόμβου προς τους υπολοίπους, αποτελώντας δείκτη της ποιότητας των οδικών υποδομών (στη σύγχρονη μορφή τους) που απολαμβάνει ο κάθε νομός της Ελλάδας κατά την επικοινωνία του με τους υπόλοιπους (Τσιώτας κά., 2012).

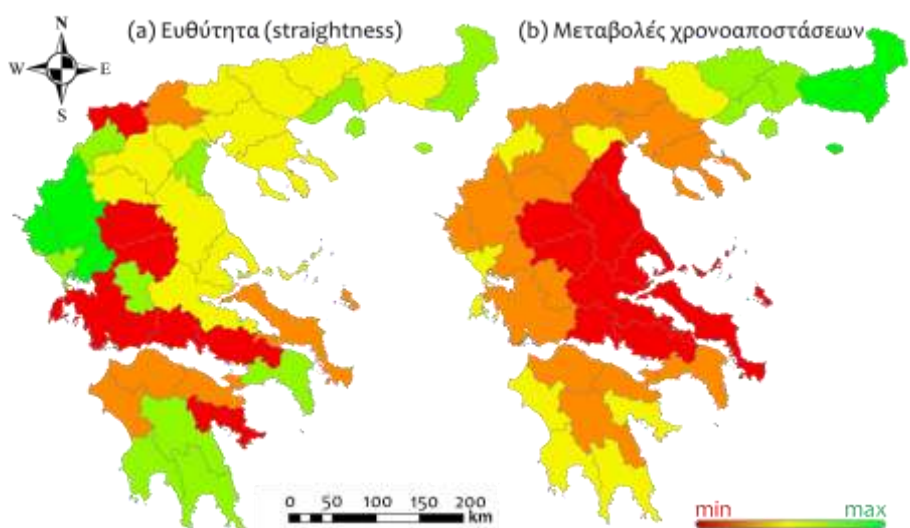

**Σχήμα 6.** Χωρική κατανομή (a) της κεντρικότητας ευθύτητας το έτος 2010 και (b) των μεταβολών των χρονοαποστάσεων ανά νομό για τις περιόδους 1988 και 2010.

Οι χρωματικές διαβαθμίσεις του σχήματος 6(a) αντιπροσωπεύουν το μέγεθος της ωφέλειας που εισέπραξαν οι νομοί της χώρας από την κατανομή των έργων υποδομής και γενικά από την πολιτική των μεταφορικών υποδομών της περιόδου 1988-2010, με κριτήριο την πρόσβασή τους στο διαπεριφερειακό δίκτυο GCN. Υπό το πρίσμα αυτό, οι νομοί που ωφελήθηκαν περισσότερο από τη χωρική κατανομή των έργων υποδομής στην Ελλάδα είναι οι *Ιωαννίνων*, *Θεσπρωτίας* και *Άρτας*. Η γεωγραφική τους θέση υποδηλώνει ότι επωφελήθηκαν τόσο από το έργο της γέφυρας *Ρίου-Αντιρίου* (που συνδέει το νομό *Αχαΐας* και *Αιτωλοακαρνανίας* παρέχοντας πρόσβαση από και προς την Πελοπόννησο από τη Δυτική Ελλάδα) όσο και της *Εγνατίας Οδού* (η οποία συνδέει το νομό *Θεσπρωτίας* με τον *Έβρου*).

Δευτερευόντως, την υψηλότερη κεντρικότητα ευθύτητας εμφανίζουν:

• Η συστάδα των νομών *Καστοριάς*, *Πρέβεζας*, *Ευρυτανίας*, οι οποίοι πρόσκεινται στους νομούς με τη μεγαλύτερη $C^s$, με αποτέλεσμα η ωφέλεια που αποκόμισαν να είναι αντίστοιχη της προαναφερόμενης περίπτωσης.

• Η συστάδα των νομών *Αρκαδίας* και *Λακωνίας* και *Μεσσηνίας* στην Πελοπόννησο, οι οποίοι προφανώς ευνοήθηκαν περισσότερο από τη ζεύξη *Ρίου-Αντιρίου*,

• Ο νομός *Αττικής*, ο οποίος ευνοήθηκε από το σύνολο σχεδόν των έργων αναβάθμισης του οδικού δικτύου, καθώς και

• Οι νομοί *Καβάλας* και *Έβρου* στη βόρεια Ελλάδα, οι οποίοι φαίνεται πως ευνοήθηκαν κατά βάση από την κατασκευή της Εγνατίας Οδού.

Έπειτα, η γεωγραφική κατανομή των της μεταβολής των χρονοαποστάσεων (σχήμα 6.b) παρουσιάζει τους νομούς που ωφελήθηκαν περισσότερο σε χρόνο ταξιδιού από την ελληνική πολιτική των μεταφορών της περιόδου 1988-2010 (Τσιώτας κά., 2012). Όπως φαίνεται στο χάρτη, η κατανομή των διαφορών των χρονοαποστάσεων εμφανίζει μία σαφή χωρική ομαδοποίηση που έχει τις μεγάλες τιμές της στην περιφέρεια και τις μικρότερες στο κέντρο. Ειδικότερα, οι νομοί που εμφάνισαν μεγαλύτερο χρονικό κέρδος στις διαπεριφερειακές τους μετακινήσεις είναι κυρίως οι μεθοριακοί νομοί του *Έβρου* και της *Ροδόπης* και δευτερευόντως οι νομοί της *Ξάνθης*, της *Καβάλας* και της *Δράμας*. Στον αμέσως επόμενο βαθμό σημαντικότητας (με κίτρινο χρώμα) οι νομοί που ευνοήθηκαν σε χρόνο διαπεριφερειακής μετακίνησης είναι οι *Σερρών*, *Ημαθίας*, και *Φλώρινας*, στη Βόρεια Ελλάδα, οι νομοί *Πρέβεζας* και *Λευκάδος* στη Δυτική και η συστάδα των νομών *Ηλείας*, *Μεσσηνίας*, *Λακωνίας* και *Αργολίδας* στην Πελοπόννησο. Προφανώς, οι θέσεις των νομών που εντάσσονται στις παραπάνω περιπτώσεις οδηγεί στην εξαγωγή πρόδηλων συμπερασμάτων σχετικά με τα έργα υποδομής (κατασκευή της Εγνατίας Οδού, αναβάθμιση της ΠΑΘΕ, ζεύξη Ρίου-Αντιρρίου) που επέδρασαν κατά περίπτωση στη μείωση των χρόνων των διαπεριφερειακών μετακινήσεων.

Στο τελευταίο στάδιο της μελέτης της τοπολογίας του GCN εξετάζονται οι συσχετίσεις των μεγεθών της *ενδιαμέσου κεντρικότητας* $C^b(k)$ και της *χωρικής ισχύος* $s(k)$ ως προς το μέγεθος του *βαθμού* $k$, με σύγκριση των διαγραμμάτων του σχήματος 7. Από την προσαρμογή καμπυλών παρεμβολής (fitting curves) στα δεδομένα διασποράς των ζευγών $(k, \langle C^b|_{k=k_i} \rangle)$ και $(k, \langle s|_{k=k_i} \rangle)$, προκύπτει η ύπαρξη αξιοσημείωτης γραμμικότητας και για τις δύο περιπτώσεις, έχοντας συντελεστές προσδιορισμού $R^2_{C^b,k}$=0.96 και $R^2_{s,k}$=0.906 αντίστοιχα.

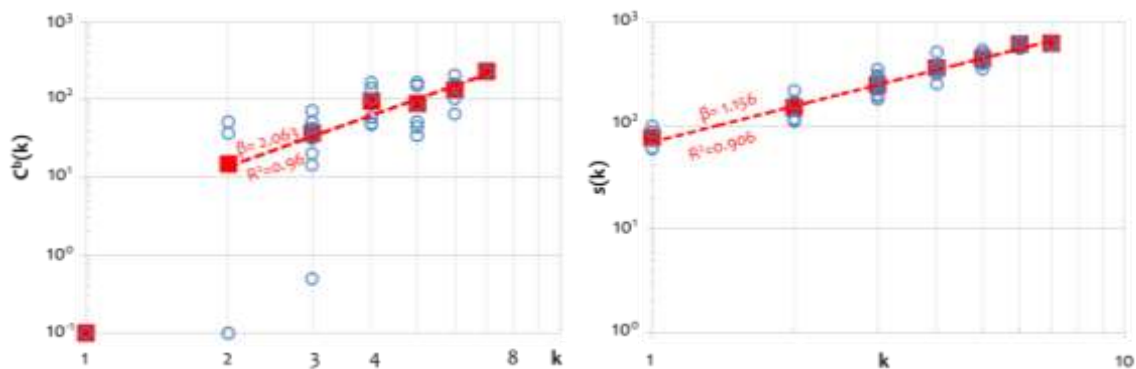

**Σχήμα 7.** Διαγράμματα διασποράς (scatter plots) (αρ.) βαθμού- ενδιαμέσου κεντρικότητας $(k,C^b)$ και (δεξ.) βαθμού-χωρικής ισχύος $(k,s)$ για το GCN. Τα κόκκινα τετράγωνα αντιστοιχούν στις μέσες τιμές για κάθε κατηγορία βαθμού.

Η σχέση $\langle C^b|_{k=k_i} \rangle = f(k)$, μεταξύ βαθμού $k$ και μέσης τιμής της ενδιαμέσου κεντρικότητας ανά βαθμό $\langle C^b|_{k=k_i} \rangle$, με $i$=2,3,…,7, έχει εκθέτη του προτύπου κανόνα δύναμης $\beta_{GCN}$=1.94 και εκφράζει ότι οι ισχυρά συνδετικοί κόμβοι στο δίκτυο (πλήμνες, hubs) αναλαμβάνουν το μεγαλύτερο φορτίο της κυκλοφορίας του. Αντίθετα, ο εκθέτης $\beta_{GCN}$=1.156 της σχέσης $\langle s|_{k=k_i} \rangle = f(k)$, μεταξύ βαθμού $k$ και μέσης ισχύος $\langle s|_{k=k_i} \rangle$, βρίσκεται κοντά στη μονάδα (~1) και υποδηλώνει μία σχετική ομοιογένεια στην ανάληψη του φορτίου των απόμακρων συνδέσεων, η οποία είναι σχεδόν γραμμική.

Τέλος, στο σχήμα 8 παρουσιάζεται η διασπορά των τιμών (*k*,*C*(*k*)), μεταξύ του συντελεστή συγκέντρωσης *C* και βαθμού *k* του GCN.

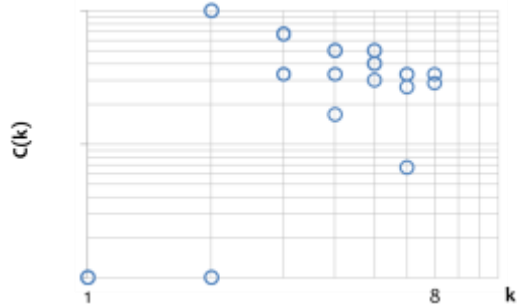

**Σχήμα 8.** Μεταβολή του συντελεστή συγκέντρωσης *C*(*k*) σε σχέση με το βαθμό κόμβων *k* του διαπεριφερειακού δικτύου commuters της Ελλάδας (GCN). Η μορφή του διαγράμματος διασποράς υποδεικνύει λογαριθμική μείωση όσο μεγαλώνουν οι τιμές του *k*.

Η σχέση *C*=*f*(*k*) (σχήμα 8) υποδεικνύει την ύπαρξη ενός μηχανισμού λογαριθμικής μείωσης της συγκέντρωσης του GCN, με την αύξηση των τιμών του *k*, η οποία είναι σύμφωνη με την κοινή ερευνητική πρακτική (Sen et al., 2003, Barthelemy, 2011). Η σχέση αυτή εκφράζει ότι όσο η συνδετικότητα ενός κόμβου αυξάνει στο δίκτυο, τόσο περιορίζεται η πιθανότητα ο κόμβος αυτός να σχετίζεται με αλληλοσυνδεδεμένους γείτονες.

**4. Εμπειρική ανάλυση**

Τα αποτελέσματα υπολογισμού των μεταβλητών που προέκυψαν ως αντιπρόσωποι των ομάδων $\mathbf{X}_S$, $\mathbf{X}_B$ και $\mathbf{X}_O$, σύμφωνα με τη μέθοδο που περιγράφηκε στην ενότητα 2.3, παρουσιάζονται στον πίνακα 5, στον οποίον η σχετική θέση (ιεραρχία) των εικονιζόμενων μεταβλητών εντός των ομάδων (within-groups calculations) παρουσιάζεται στην πρώτη στήλη, ενώ η ως προς στο σύνολο των διαθέσιμων μεταβλητών (global calculations) παρουσιάζεται στις στήλες με την ένδειξη "*Rank*".

**Πίνακας 5.**
Αποτελέσματα της ανάλυσης για την εκλογή των αντιπροσωπευτικών μεταβλητών του υποδείγματος

| Θέση μέσα στην ομάδα | **Δομικές Μεταβλητές – $\mathbf{X}_S$** | | | | **Λειτουργικές Μεταβλητές – $\mathbf{X}_B$** | | | | **Οντολογικές Μεταβλητές – $\mathbf{X}_O$** | | | |
|---|---|---|---|---|---|---|---|---|---|---|---|---|
| | Μέσα στην ομάδα (within-groups) | | Συνολικά (global) | | Μέσα στην ομάδα (within-groups) | | Συνολικά (global) | | Μέσα στην ομάδα (within-groups) | | Συνολικά (global) | |
| | Σύμβ.[a] | $\Sigma_S(r^2)$ [b] | Rank[c] | $\Sigma(r^2)$ | Συμβ. | $\Sigma_B(r^2)$ | Rank | $\Sigma(r^2)$ | Συμβ. | $\Sigma_O(r^2)$ | Rank | $\Sigma(r^2)$ |
| *1* | **$S_6$** | **2,746** | **2** | **12,116** | $Y$[d] | 5,285 | 1 | 12,181 | **$O_2$** | **4,091** | 9 | 10,451 |
| *2* | $S_5$ | 2,738 | 3 | 12,058 | **$B_6$** | **5,190** | **4** | **11,905** | $O_6$ | 3,990 | 7 | 11,703 |
| *3* | $S_{10}$ | 2,493 | 11 | 7,877 | $B_7$ | 5,152 | 6 | 11,730 | $O_7$ | 3,983 | **5** | **11,805** |
| *4* | $S_1$ | 2,448 | 23 | 2,596 | $B_8$ | 5,048 | 8 | 11,428 | $O_4$ | 3,256 | 13 | 6,443 |
| *5* | $S_2$ | 2,188 | 22 | 2,945 | $B_3$ | 3,910 | 10 | 9,400 | $O_9$ | 2,612 | 19 | 3,859 |
| *6* | $S_9$ | 1,777 | 25 | 2,137 | $B_5$ | 3,362 | 12 | 7,177 | $O_1$ | 2,507 | 14 | 5,169 |
| *7* | $S_4$ | 1,753 | 27 | 1,838 | $B_4$ | 2,622 | 16 | 4,883 | $O_3$ | 2,236 | 21 | 3,307 |
| *8* | $S_{11}$ | 1,576 | 17 | 4,404 | $B_1$ | 1,628 | 20 | 3,467 | $O_5$ | 1,935 | 18 | 3,898 |
| *9* | $S_8$ | 1,570 | 28 | 1,790 | $B_2$ | 1,000 | 30 | 1,206 | $O_8$ | 1,849 | 15 | 5,080 |
| *10* | $S_3$ | 1,569 | 24 | 2,169 | | | | | $O_{10}$ | 1,368 | 29 | 1,458 |
| *11* | $S_7$ | 1,158 | 26 | 1,931 | | | | | | | | |

a. Σύμβολο μεταβλητής (βλ. αναλυτική περιγραφή στον πίνακα 10)
b. Άθροισμα τετραγώνων των συντελεστών συσχέτισης
c. Σχετική θέση του μεγέθους της κάθε μεταβλητής ως προς το σύνολο των υπολοίπων (global calculations)

| Θέση μέσα στην ομάδα | Δομικές Μεταβλητές – $X_S$ | | | | Λειτουργικές Μεταβλητές – $X_B$ | | | | Οντολογικές Μεταβλητές – $X_O$ | | | |
|---|---|---|---|---|---|---|---|---|---|---|---|---|
| | Μέσα στην ομάδα (within-groups) | | Συνολικά (global) | | Μέσα στην ομάδα (within-groups) | | Συνολικά (global) | | Μέσα στην ομάδα (within-groups) | | Συνολικά (global) | |
| | Σύμβ.[a] | $\Sigma_S(r^2)$[b] | Rank[c] | $\Sigma(r^2)$ | Συμβ. | $\Sigma_B(r^2)$ | Rank | $\Sigma(r^2)$ | Συμβ. | $\Sigma_O(r^2)$ | Rank | $\Sigma(r^2)$ |

d. Η μεταβλητή αυτή εξαιρείται από την ανάλυση γιατί χρησιμοποιείται στο υπόδειγμα ως εξαρτημένη μεταβλητή

Σύμφωνα με τα αποτελέσματα του πίνακα 5, οι αντιπροσωπευτικές μεταβλητές που προκύπτουν από την ανάλυση μέσα στις κατηγορίες (within-groups calculations) είναι η μεταβλητή $S_6$ (πληθυσμός) για την ομάδα *των δομικών μεταβλητών*, η $B_6$ (αριθμός ΙΧ οχημάτων) για την κατηγορία *των λειτουργικών μεταβλητών* και η $O_2$ (δείκτης εκπαίδευσης) για την *οντολογική* ομάδα. Τα αντίστοιχα αποτελέσματα για τη συνολική περίπτωση (global analysis) διαφέρουν ελαφρώς αναδεικνύοντας ως αντιπροσώπους τις μεταβλητές $S_6$, $B_6$ και $O_7$ (αριθμός ατυχημάτων) αντί για $O_2$, αντίστοιχα. Λαμβάνοντας υπόψη ότι η θέση της μεταβλητής $O_2$ για την περίπτωση εντός των ομάδων (within-class analysis) είναι μετατοπισμένη τρεις θέσεις χαμηλότερα σε σχέση με τη συνολική περίπτωση (global analysis), για χάρη πληρότητας εξετάζονται οι τρεις (αντί της μίας) ομάδες αντιπροσώπων ($S_6$, $B_6$, $O_7$), ($S_6$, $B_6$, $O_7$) και ($S_6$, $B_6$, $O_7$), συνυπολογίζοντας δηλαδή τις μεταβλητές της οντολογικής ομάδας που υπερτερούν στη συνολική περίπτωση από την $O_2$.

Μια ενδιαφέρουσα παρατήρηση που προκύπτει από τα αποτελέσματα του πίνακα 5 αφορά το γεγονός ότι η εξαρτημένη μεταβλητή $Y$ (αριθμός commuters) τοποθετείται και στις δύο περιπτώσεις (within-class και global) της ανάλυσης στην πρώτη θέση της ταξινόμησης, εμφανίζοντας τα μεγαλύτερα αθροίσματα τετραγώνων των συντελεστών συσχέτισης. Το αποτέλεσμα αυτό φαίνεται λογικό, δεδομένου ότι οι διαθέσιμες ανεξάρτητες μεταβλητές έχουν επιλεχθεί με κριτήριο την άμεση ή έμμεση συνάφειά τους με το φαινόμενο του commuting, με αποτέλεσμα πολλές από αυτές να αποτελούν προσδιοριστικούς παράγοντες της ημερήσιας μετακίνησης με σκοπό την εργασία. Επίσης ενδιαφέρουσα παρατήρηση αποτελεί το γεγονός ότι η μεταβλητή του πληθυσμού ($S_6$) εμφανίζει τη μεγαλύτερη συσχέτιση με το σύνολο των υπόλοιπων ανεξάρτητων μεταβλητών, προσδίδοντας στο φαινόμενο του commuting ένα βαρυτικό χαρακτήρα (Tsiotas and Polyzos, 2015b).

Στο τελευταίο στάδιο οι μεταβλητές που επιλέγονται από την παραπάνω διαδικασία ως αντιπρόσωποι των κατηγοριών $X_S$, $X_B$ και $X_O$ εισάγονται ως ανεξάρτητες μεταβλητές ($X_i$) σε ένα υπόδειγμα πολυμεταβλητής γραμμικής παλινδρόμησης, με εξαρτημένη μεταβλητή $Y$ τον αριθμό των ημερησίως μετακινουμένων με σκοπό την εργασία (commuters). Στον πίνακα 6 παρουσιάζονται τα αποτελέσματα της πολυμεταβλητής ανάλυσης γραμμικής παλινδρόμησης, η οποία εφαρμόζεται στα τρία διαφορετικά σύνολα των αντιπροσώπων μεταβλητών $Y_1=(S_6,B_6,O_2)$, $Y_2=(S_6,B_6,O_7)$ και $Y_3=(S_6,B_6,O_6)$ που αναδείχθηκαν στο προηγούμενο στάδιο της ανάλυσης.

**Πίνακας 6**
Αποτελέσματα της πολυμεταβλητής ανάλυσης γραμμικής παλινδρόμησης

| Υπόδειγμα[a] | | Ανεξ. μεταβλητές | Μη τυποποιημένοι συντελεστές | | Τυπ. συντελ | | |
|---|---|---|---|---|---|---|---|
| Πληροφορίες υποδείγματος | | | $b$[b] | Τυπ. σφάλμα | $b$[c] | $t$[d] | Sig.[e] |
| ($Y_1$) Ανεξάρτητες μεταβλητές υποδείγματος: (Σταθερά), $S_6$, $B_6$, $O_2$ | | | | | | | |
| $R$[f] | 0,999 | (Σταθερά) | -881,91 | 174,10 | | -5,066 | 0,000 |
| $R^{2}$[g] | 0,998 | $S_6$ | 0,011 | 0,002 | 0,600 | 4,840 | 0,000 |
| Τυπ. σφάλμα εκτίμησης | 600,75 | $B_6$ | 0,011 | 0,004 | 0,345 | 3,003 | 0,005 |
| | | $O_2$ | 41,422 | 11,729 | 0,065 | 3,532 | 0,001 |

| Υπόδειγμα[a] | | | Μη τυποποιημένοι συντελεστές | | Τυπ. συντελ | | |
|---|---|---|---|---|---|---|---|
| Πληροφορίες υποδείγματος | | Ανεξ. μεταβλητές | $b$[b] | Τυπ. σφάλμα | $b$[c] | $t$[d] | Sig.[e] |
| ($Y_2$) Ανεξάρτητες μεταβλητές υποδείγματος: (Σταθερά), $S_6$, $B_6$, $O_7$ | | | | | | | |
| $R$ | 0,998 | (Σταθερά) | -674,49 | 212,47 | | -3,174 | 0,003 |
| $R^2$ | 0,997 | $S_6$ | 0,015 | 0,002 | 0,816 | 6,151 | 0,000 |
| Τυπ. σφάλμα εκτίμησης | 693,20 | $B_6$ | 0,02 | 0,005 | 0,049 | 0,335 | 0,740 |
| | | $O_7$ | 1,198 | 1,482 | 0,134 | 0,809 | 0,424 |
| ($Y_3$) Ανεξάρτητες μεταβλητές υποδείγματος: (Σταθερά), $S_6$, $B_6$, $O_6$ | | | | | | | |
| $R$ | 0,999 | (Σταθερά) | -532,81 | 199,46 | | -2,671 | 0,011 |
| $R^2$ | 0,997 | $S_6$ | 0,016 | 0,002 | 0,834 | 7,991 | 0,000 |
| Τυπ. σφάλμα εκτίμησης | 643,41 | $B_6$ | 0,021 | 0,007 | 0,627 | 2,796 | 0,008 |
| | | $O_6$ | -6,337 | 2,508 | -0,462 | -2,527 | 0,016 |

a. Μέθοδος Enter (υπολογισμός του συνόλου των μεταβλητών)
b. Μη τυποποιημένοι συντελεστές *βήτα* του υποδείγματος
c. Τυποποιημένοι συντελεστές *βήτα* του υποδείγματος
d. Στατιστικό *t* για τον έλεγχο της σημαντικότητας των συντελεστών
e. Δίπλευρη σημαντικότητα (2-tailed)
f. Πολλαπλός συντελεστής συσχέτισης
g. Συντελεστής προσδιορισμού

Οι τιμές των συντελεστών προσδιορισμού ($R^2$) στον πίνακα 6 εκφράζουν ότι τα τρία υποδείγματα $Y_1$, $Y_2$ και $Y_3$ διακρίνονται από σχεδόν άριστη ικανότητα περιγραφής της μεταβλητότητας των τιμών της εξαρτημένης μεταβλητής (αριθμός των μετακινουμένων με σκοπό την εργασία), με βάση τη μεταβλητότητα των τιμών των μεταβλητών πρόβλεψης στο κάθε πρότυπο. Η πολύ καλή ικανότητα προσδιορισμού των υποδειγμάτων συνάγεται και από την άθροιση των τυποποιημένων συντελεστών της παλινδρόμησης, η οποία σε κάθε περίπτωση αγγίζει τη μονάδα, υποδηλώνοντας την παρουσία αμελητέας συγγραμικότητας (collinearity) μεταξύ των μεταβλητών (Tsiotas and Polyzos, 2015a). Το παραπάνω γεγονός επικυρώνει τη χρησιμότητα της μεθοδολογικής προσέγγισης που επιχειρήθηκε για τη μελέτη του διαπεριφερειακού δικτύου των commuters, ιδιαίτερα εφόσον ληφθεί υπόψη ότι μόνο το ~10% της διαθέσιμης πληροφορίας (3 από τις 29 μεταβλητές) χρησιμοποιήθηκε για την κατασκευή των υποδειγμάτων.

Η διαδικασία επιλογής των αντιπροσωπευτικών μεταβλητών προσδίδει μια συστολική ιδιότητα στη μεθοδολογία κατασκευής των υποδειγμάτων της γραμμικής παλινδρόμησης, προσδιορίζοντάς τα, ταυτόχρονα, εντός του εννοιολογικού πλαισίου της Επιστήμης των Δικτύων. Ενδιαφέρουσα προοπτική για περαιτέρω έρευνα αποτελεί η σύγκριση των αποτελεσμάτων με άλλες μεθόδους, όπως η *ανάλυση σε κύριες συνιστώσες* (*principal component analysis*) (Norusis, 2004) ή η μεθόδος των αποβολών (Norusis, 2004; Tsiotas and Polyzos, 2015a,b).

Συνολικά, τα αποτελέσματα του πίνακα 6 υποδηλώνουν ότι η μεταβλητή του πληθυσμού ($S_6$) αποτελεί το σημαντικότερο προσδιοριστικό παράγοντα στην περιγραφή του φαινομένου commuting στην Ελλάδα. Η παρατήρηση αυτή συμφωνεί με τη θεωρία (Πολύζος, 2011; Polyzos et al., 2014, 2015), αναδεικνύοντας τη βαρυτική διάσταση της ημερήσιας μετακίνησης με σκοπό την εργασία. Ειδικότερα, το ποσοστό της συμβολής της μεταβλητής $S_6$ στο υπόδειγμα, όπως αυτό συνάγεται από τις τιμές των τυποποιημένων συντελεστών *βήτα*, κυμαίνεται μεταξύ 60-83%. Περαιτέρω, η παρουσία της μεταβλητής του αριθμού των ΙΧ ($B_6$) στα υποδείγματα υποδηλώνει πως η χρήση ιδιόκτητων οχημάτων από τους εργαζόμενους διαδραματίζει καθοριστικό ρόλο στη διαμόρφωση του φαινομένου της ημερήσιας διαπεριφερειακής μετακίνησης με σκοπό την εργασία. Στην κλίμακα των διαπεριφερειακών μετακινήσεων φαίνεται πως η χρήση των εναλλακτικών τρόπων μεταφοράς (λεωφορείο, τραίνο) δεν αποτελεί πρώτιστη επιλογή για την ημερήσια μετακίνηση με σκοπό την εργασία, προφανώς λόγω του γεγονότος ότι τα χρονικά μεγέθη της μετακίνησης καθίστανται ήδη ιδιαίτερα κρίσιμα, ώστε να μην επιδέχονται την περαιτέρω χρονική επιβάρυνση από τη μετακίνηση προς και από τους σταθμούς, αλλά ούτε και από την αναμονή στο σταθμό μέχρι την αναχώρηση του μέσου μεταφοράς.

Για το υπόδειγμα $Y_1$ η συνεισφορά της μεταβλητής $B_6$ προσεγγίζεται στο 34,5%, λαμβάνοντας υπόψη ότι η δράση της μεταβλητής του επιπέδου εκπαίδευσης των εργαζομένων ($O_2$) περιορίζεται στο 6,5%. Στο υπόδειγμα $Y_2$ τόσο η συνεισφορά της μεταβλητής $B_6$ όσο και της αντίστοιχης $O_7$ (αριθμός τροχαίων ατυχημάτων) θεωρείται στατιστικά ασήμαντη, εκφράζοντας πιθανώς ότι ο αριθμός των ΙΧ οχημάτων ($B_6$) (στο βαθμό που συνδέεται με το ενδεχόμενο πρόκλησης ατυχήματος – $O_7$), δεν αποτελούν κριτήρια διεξαγωγής του φαινομένου commuting.

Τέλος, για το υπόδειγμα $Y_1$ η συνεισφορά της μεταβλητής $B_6$ προσεγγίζεται στο 62,7%, αλλά μαζί με τη μεταβλητή του πληθυσμού $S_6$ δρουν ανταγωνιστικά στη μεταβλητή $S_6$ που εκφράζει το προϊόν του νομού στον τομέα μεταφορών. Η εικόνα που σκιαγραφείται από το πρότυπο αυτό προφανώς αναδεικνύει μία βαρυτική πτυχή του commuting, καθόσον φαίνεται πως περιγράφει την ελκτική δράση των πολυπληθών πόλεων, οι οποίες διατηρούν ένα μεγάλο ποσό δραστηριότητας commuting και αριθμού εργαζομένων εντός των αστικών τους ορίων, περιορίζοντας έτσι τις διαπεριφερειακές μεταφορές και συνεπώς το παραγόμενο από αυτές προϊόν.

## 5. Συμπεράσματα

Στο άρθρο αυτό μελετήθηκε η τοπολογία του *διαπεριφερειακού δικτύου των ημερησίως μετακινουμένων με σκοπό την εργασία* (GCN). Στη μελέτη επιδιώχθηκε η εξόρυξη της κοινωνικοοικονομικής πληροφορίας που είναι ενσωματωμένη στην τοπολογία αυτού του δικτύου, υπό το πρίσμα της Ανάλυσης των Σύνθετων Δικτύων και της Στατιστικής Μηχανικής, με σκοπό τη διερεύνηση των παραμέτρων που διαμορφώνουν το πλαίσιο λειτουργίας των ημερήσιων διαπεριφερειακών μετακινήσεων στην Ελλάδα. Το GCN αναπαραστάθηκε στον *L-χώρο* αντιπροσώπευσης ως μη κατευθυνόμενος γράφος, του οποίου οι κόμβοι αντιπροσωπεύουν τις *πρωτεύουσες* των ελληνικών νομών, ενώ οι ακμές *την ύπαρξη δυνατότητας απευθείας οδικών συνδέσεων* μεταξύ των νομών της Ελλάδας.

Στην ανάλυση που πραγματοποιήθηκε υπήρξε εμφανής η επίδραση των χωρικών περιορισμών, όπως προέκυψε από την συνδυασμένη ερμηνεία των παρακάτω παρατηρήσεων:

• από το οξυμένο πρότυπο στην κατανομή του βαθμού (peaked degree distribution), τα οποία αποκλίνουν από την περιγραφή του προτύπου κανόνα-δύναμης (power-law) που χαρακτηρίζει περιπτώσεις δικτύων με ηπιότερους χωρικούς περιορισμούς.

• μέσα από τη σημαντική συγκέντρωση των τιμών γύρω από την κύρια διαγώνιο στα πρότυπα σποραδικότητας (sparsity patterns – spy plots) του πίνακα συνδέσεων, η οποία ανέδειξε την ύπαρξη χαρακτηριστικών *δικτυώματος* (*lattice network*).

• μέσα από τον υπολογισμό του ω-δείκτη, ο οποίος χρησιμοποιείται για την προσεγγιστική ανίχνευση της τυπολογίας των δικτύων και κυρίως του *μικρού-κόσμου* (*small-world property*) και που περιέγραψε την ύπαρξη ιδιοτήτων δικτυώματος για το GCN.

• με την εμφάνιση των κεντρικών γεωγραφικών θέσεων του δικτύου να πλεονεκτούν στην κατανομή των μεγεθών του βαθμού και της ενδιαμέσου κεντρικότητας.

• με τον επιμερισμό του GCN σε κοινότητες γεωγραφικής συνάφειας (modularity optimization) και

• με τα πρότυπα *κανόνα-δύναμης* (*power-law*) που διαμορφώθηκαν κατά τις συσχετίσεις του μεγέθους του βαθμού ($k$) με την ενδιαμέσου κεντρικότητα ($C^b$), τη χωρική ισχύ ($s$) και το συντελεστή συγκέντρωσης ($C$).

- από τις μεγάλες διακυμάνσεις που εμφανίστηκαν στο μέγεθος της στην ενδιαμέσου κεντρικότητας ($C^b$), οι οποίες υποδηλώνουν ότι το μέγεθος αυτό αποκτά σαφή γεωγραφική υπόσταση, τείνοντας να ταυτιστεί με την έννοια του κέντρου βάρους των κόμβων του δικτύου.

Παρά την καθοριστική επίδραση των χωρικών περιορισμών στη διαμόρφωση της τοπολογίας του GCN, η μορφή της σχέσης $s=f(k)$ έδειξε την ύπαρξη συνδέσεων μεγάλου μήκους, οι οποίες ανιχνεύονται στην περίπτωση που ο εκθέτης του προτύπου κανόνα-δύναμης (power-law) είναι μεγαλύτερος της μονάδας ($β >1$) (Barthelemy, 2011). Το γεγονός αυτό οφείλεται προφανώς στον κανόνα δημιουργίας του δικτύου, του οποίου οι ακμές έχουν εννοιολογική (εκφράζουν τη λειτουργία και τις σχέσεις οδικής επικοινωνίας που αναπτύσσονται στο διοικητικό πλέγμα των νομών της Ελλάδας) και όχι φυσική σημασία.

Περαιτέρω, μέσα από την αντιπαραβολή των μέτρων κεντρικότητας που υπολογίστηκαν για δύο διαφορετικές χρονικές εκφάνσεις (1988 και 2010) του GCN εξήχθηκε πληροφορία σχετικά με τα μεγαλύτερα έργα υποδομής που συντελέστηκαν στον τομέα των οδικών μεταφορών και επηρέασαν τη μεταφορική ικανότητα της χώρας. Οι αλλαγές που εντοπίστηκαν στην κεντρικότητα των νομών την περίοδο 1988-2010 παρέχουν υλικό για την αξιολόγηση της πολιτικής που ασκήθηκε στον τομέα των μεταφορικών υποδομών, καθόσον αποκαλύφθηκαν οι νομοί που επωφελήθηκαν περισσότερο από τα έργα υποδομής της περιόδου 1988-2010. Σύμφωνα με την ανάλυση που πραγματοποιήθηκε, η πολιτική της Ελλάδας στις υποδομές των μεταφορών της περιόδου 1988-2010 εμφανίζεται να κατευθύνεται κάτω από ένα στοχοθετημένο προγραμματισμό και να παρουσιάζεται ευνοϊκή στις μεθόριες και απομακρυσμένες περιοχές, επιδιώκοντας την εξάλειψη των γεωγραφικών ανισοτήτων και τη δημιουργία αναπτυξιακών προοπτικών.

Στο μέρος της εμπειρικής ανάλυσης, πραγματοποιήθηκε κατασκευή ενός (προτύπου) υποδείγματος πολυμεταβλητής γραμμικής παλινδρόμησης, το οποίο διαρθρώθηκε πάνω στις τρεις εννοιολογικές συνιστώσες που προτάθηκαν για τη μελέτη των χωρικών δικτύων από τους Tsiotas and Polyzos (2015c). Στην ανάλυση συμμετείχαν 30 διανυσματικές μεταβλητές που περιλαμβάνουν τις τιμές των νομών για κάθε χαρακτηριστικό, οι οποίες ομαδοποιήθηκαν ως προς τη θεματική τους συνάφεια. Στη συνέχεια, επιλέχθηκαν τρεις ομάδες τριών αντιπροσώπων από κάθε κατηγορία, με τη μεγαλύτερη συσχέτιση μέσα σε κάθε ομάδα και ξεχωριστά, οι οποίοι χρησιμοποιήθηκαν ως ανεξάρτητες μεταβλητές (μεταβλητές πρόβλεψης) για την κατασκευή ισάριθμων υποδειγμάτων παλινδρόμησης. Τα τρία υποδείγματα $Y_1$, $Y_2$ και $Y_3$ που κατασκευάστηκαν εμφάνισαν άριστη ικανότητα προσδιορισμού και αμελητέα συγγραμμικότητα. Τα αποτελέσματα της ανάλυσης ανέδειξαν τη βαρυτική διάσταση της ημερήσιας μετακίνησης με σκοπό την εργασία, μέσα από τη συμμετοχή της μεταβλητής του πληθυσμού στα τρία υποδείγματα σε ποσοστό 60-83%.

Τέλος, από την παρουσία των υπολοίπων μεταβλητών στο υπόδειγμα διαφάνηκε ο καθοριστικός ρόλος που διαδραματίζει η χρήση ιδιόκτητων οχημάτων από τους εργαζόμενους στη διαμόρφωση του φαινομένου της ημερήσιας διαπεριφερειακής μετακίνησης με σκοπό την εργασία, η συνεισφορά του οποίου κυμαίνεται περίπου από 5-63%. Στο δίκτυο των διαπεριφερειακών μετακινήσεων διαφάνηκε πως η χρήση των εναλλακτικών τρόπων μεταφοράς (λεωφορείο, τραίνο) δεν αποτελεί πρώτιστη επιλογή για την ημερήσια μετακίνηση με σκοπό την εργασία, ενώ το επίπεδο εκπαίδευσης των εργαζομένων αποτελεί κίνητρο που επηρεάζει το φαινόμενο σε ποσοστό 6.5%. Τέλος, η ανάλυση σκιαγράφησε την ελκτική δράση των πολυπληθών πόλεων, οι οποίες διατηρούν ένα μεγάλο ποσό δραστηριότητας commuting και αριθμού εργαζομένων εντός των αστικών τους ορίων, περιορίζοντας έτσι τις διαπεριφερειακές μεταφορές και συνεπώς το παραγόμενο από αυτές προϊόν.

## 6. Βιβλιογραφία

Ozbay, K., Bartin, B., Yanmaz-Tuzel, O., Berechman, J., (2007) "Alternative methods for estimating full marginal costs of highway transportation", *Transportation Research Part A*, 41, pp.768-786.

## 7. Παράρτημα

**Πίνακας 1**

Μέτρα χώρου και τοπολογίας που χρησιμοποιούνται στην ανάλυση του GRN

| Μέτρο[*] | Περιγραφή | Μαθηματική Έκφραση | Αναφορά |
|---|---|---|---|
| *Πυκνότητα γράφου - Graph density* ($\rho$) | Ο λόγος του αριθμού των υφιστάμενων συνδέσεων (ακμών) του δικτύου προς τον αριθμό των δυνατών συνδέσεων που μπορούν να σχηματιστούν από το σύνολο των κόμβων. Το μέγεθος της πυκνότητας αντιπροσωπεύει την πιθανότητα εμφάνισης μιας σύνδεσης μεταξύ δύο τυχαίων κόμβων στο δίκτυο. | $\rho = \dfrac{|E(G)|}{|E(G_{complete})|} =$ $= m \Big/ \binom{n}{2} = \dfrac{2m}{n \cdot (n-1)}$ | (Tsiotas and Polyzos, 2015a) |
| *Βαθμός κόμβου - Node Degree* ($k$) | Ο αριθμός των προσκείμενων ακμών σε μία κορυφή του δικτύου, ο οποίος αντιπροσωπεύει τη συνδετικότητα και την ικανότητα επικοινωνίας του δικτύου. | $k_i = k(i) = \sum_{j \in V(G)} \delta_{ij}$, $\delta_{ij} = \begin{cases} 1, & e_{ij} \in E(G) \\ 0, & \text{διαφορετικά} \end{cases}$ | (Koschutzki et al., 2005) |
| *Χωρική ισχύς - Node (spatial) strength* ($s$) | Το άθροισμα των χωρικών αποστάσεων των ακμών που πρόσκεινται σε έναν κόμβο. | $s_i = s(i) = \sum_{j \in V(G)} d_{ij}$, $\delta_{ij} = \begin{cases} 1, & e_{ij} \in E(G) \\ 0, & \text{otherwise} \end{cases}$ | (Barthelemy, 2011) |
| *Μέσος βαθμός κόμβων - Average Network's Degree* $\langle k \rangle$ | Ο μέσος όρος των τιμών του βαθμού των κόμβων ($k_i$) για το σύνολο των κορυφών $V(G)$ του δικτύου. | $\langle k \rangle = \dfrac{1}{|V(G)|} \cdot \sum_{i=1}^{|V(G)|} k(i) =$ $= \dfrac{1}{n} \cdot \sum_{i=1}^{n} k(i)$ | (Barthelemy, 2011) |
| *Κεντρικότητα εγγύτητας - Closeness Centrality*[*] ($C_i^C$) | Ισούται με το αντίστροφο μέσο μήκος των ελάχιστων μονοπατιών που ξεκινούν από έναν δεδομένο κόμβο $v \in V(G)$ και εκφράζει την προσβασιμότητα του κόμβου αυτού προς τους υπόλοιπους κόμβους του δικτύου. | $C_i^c = \dfrac{|V|-1}{\sum_{j=1, i \neq j}^{|V|} d_{ij}} =$ $= (\bar{d}_i)^{-1}$ | (Koschutzki et al., 2005; Tsiotas and Polyzos, 2013a). |
| *Ενδιαμέσου κεντρικότητα - Betweenness Centrality*[*] ($C_k^B$) | Ισούται με το λόγο του αριθμού των ελάχιστων μονοπατιών $\sigma(k)$ του δικτύου, τα οποία περιλαμβάνουν μία δεδομένη κορυφή k, προς το συνολικό αριθμό σ των μονοπατιών του δικτύου. | $C_k^b = \sigma(k)/\sigma$ | (Koschutzki et al., 2005) |
| *Συντελεστής συγκέντρωσης - Clustering Coefficient* ($C_v$) | Εκφράζει την πιθανότητα εύρεσης συνδεδεμένων γειτόνων σε έναν τυχαίο κόμβο του δικτύου, η οποία ισοδυναμεί με το λόγο του αριθμού των συνδεδεμένων γειτόνων $E(v)$ της κορυφής, προς τον αριθμό των συνολικών τριπλετών που σχηματίζονται από τη συγκεκριμένη κορυφή. | $c_v = \dfrac{\tau \rho i \gamma \omega \nu \alpha(v)}{\tau \rho i \pi \lambda \epsilon \tau \epsilon \varsigma(v)} =$ $= \dfrac{E(v)}{k_v \cdot (k_v - 1)}$ | (Barthelemy, 2011; Tsiotas and Polyzos, 2015a) |
| *Συναρμολογησι μότητα - Modularity* ($Q$) | Αντικειμενική συνάρτηση που εκφράζει τη δυνατότητα διαχωρισμού του δικτύου σε κοινότητες, όπου το $g_i$ αντιπροσωπεύει την κοινότητα του | $Q = \dfrac{\sum_{i,j} [A_{ij} - P_{ij}] \cdot \delta(g_i, g_j)}{2m}$ | (Blondel et al., 2008; Fortunato, 2010) |

| Μέτρο[*] | Περιγραφή | Μαθηματική Έκφραση | Αναφορά |
|---|---|---|---|
| | κόμβου $v_i$, το $[A_{ij} - P_{ij}]$ τη διαφορά του παρατηρούμενου μείον τον αναμενόμενο αριθμό των ακμών που προσπίπτουν σε ένα δεδομένο ζεύγος κορυφών $v_i,v_j$ του δικτύου και $δ(g_i,g_j)$ είναι η δείκτρια συνάρτηση που επιστρέφει την τιμή 1 όταν $g_i=g_j$. | | |
| Μέσο μήκος μονοπατιού - Average Path Length $\langle l \rangle$ | Η μέση τιμή του ελάχιστου αριθμού των ακμών $d(v_i,v_j)$ που παρεμβάλλονται για τη σύνδεση δύο τυχαίων κορυφών του δικτύου. | $\langle l \rangle = \dfrac{\sum_{v \in V(G)} d(v_i,v_j)}{n \cdot (n-1)}$ | (Barthelemy, 2011) |

\* Όταν το μέγεθος υπολογίζεται σε δυαδικές (τοπολογικές) αποστάσεις θεωρείται *δυαδικό* (*binary* measure) και συμβολίζεται με το δείκτη *bin*, ενώ όταν υπολογίζεται σε χωρικές αποστάσεις (μετρούμενες σε ναυτικά μίλια) θεωρείται χωρικά σταθμισμένο (*weighted* measure) και συμβολίζεται με το δείκτη *wei*)
(πηγή: ίδια επεξεργασία)

**Πίνακας 2**
Περιγραφή των διανυσματικών μεταβλητών που συμμετέχουν στην εμπειρική ανάλυση του GCN

| Μεταβλητή* | Περιγραφή | Πηγή |
|---|---|---|
| *Ομάδα Δομικών Μεταβλητών* (*Structural Class*) - $\mathbf{X}_S$ | | |
| ($S_1$) Βαθμός οδικού δικτύου | Ο αριθμός των συνδέσεων κάθε κόμβου του διαπεριφερειακού δικτύου commuting. | (Tsiotas and Polyzos, 2015c; Google Maps, 2013) |
| ($S_2$) Βαθμός Commuting | Ο αριθμός των προορισμών commuting που έχει κάθε κόμβος του GCN. | (Tsiotas and Polyzos, 2015c; ΕΣΥΕ, 2007) |
| ($S_3$) Διαφορά βαθμού | Η διαφορά $S_3 = S_1 - S_2$ | (Tsiotas and Polyzos, 2015c) |
| ($S_4$) Κεντρικότητα εγγύτητας (closeness centrality) | Δείκτης προσβασιμότητας που είναι αντιστρόφως ανάλογος της μέσης απόστασης μιας πόλης προς τις πόλεις του υπολοίπου του δικτύου. | (Tsiotas et al., 2013c) |
| ($S_5$) Κεντρικότητα κινητικότητας (mobility centrality) | Δείκτης κεντρικότητας, ο οποίος αποτελεί ανάλογο της κινητικής ενέργειας σωματιδίου και μετρά το δυναμικό που μία ιδιότητα κόμβου προκαλεί στο δίκτυο. | (Tsiotas and Polyzos, 2013a, 2015a) |
| ($S_6$) Πληθυσμός | Ο πληθυσμός των νομών με βάση την απογραφή του 2011. | (Tsiotas and Polyzos, 2015a,b,c) |
| ($S_7$) Πρόσημο commuting | Τριχοτομική μεταβλητή που προκύπτει από τη στατιστική διαφορά του αριθμού των εξερχομένων μείον των εισερχομένων commuters ανά πόλη και που υποδηλώνει το ρόλο της κάθε πόλης στο πρότυπο του GCN (+: απωστικός, 0: ουδέτερος, -: ελκτικός) | (Tsiotas and Polyzos, 2015c) |
| ($S_8$) Ελάχιστη απόσταση commuting | Η απόσταση του εγγύτερου προορισμού commuting για κάθε πόλη. | (Tsiotas and Polyzos, 2015c) |
| ($S_9$) Μέση απόσταση γειτόνων | Η μέση απόσταση μιας πόλης από τους commuting προορισμούς της. | (Tsiotas and Polyzos, 2015c) |
| ($S_{10}$) Αριθμός υπεραστικών προορισμών ΚΤΕΛ | Το πλήθος των προορισμών των δρομολογίων του υπεραστικού ΚΤΕΛ για κάθε πόλη. | (Tsiotas and Polyzos, 2015c) |
| ($S_{11}$) Αριθμός προορισμών ΟΣΕ | Το πλήθος των προορισμών των δρομολογίων ΟΣΕ για κάθε πόλη. | (Tsiotas and Polyzos, 2015c) |
| *Ομάδα Λειτουργικών Μεταβλητών* (*Functional Class*) - $\mathbf{X}_B$ | | |
| ($Y$)** Αριθμός commuters | Ο αριθμός των ημερησίως μετακινουμένων με σκοπό την εργασία (commuters) για κάθε | (ΕΣΥΕ, 2007) |

| Μεταβλητή* | Περιγραφή | Πηγή |
|---|---|---|
| | πόλη προορισμού. | |
| ($B_1$) Κατευθυνόμενοι commuters | Σύνθετη μεταβλητή που προκύπτει από τη σχέση: $B_1$=max{incoming, outgoing commuters}·$S_7$. | (Tsiotas and Polyzos, 2015c) |
| ($B_2$) Μέση συχνότητα υπεραστικών δρομολογίων ΚΤΕΛ | Ο μέσος όρος του αριθμού των εβδομαδιαίων δρομολογίων υπεραστικού ΚΤΕΛ για το σύνολο των δυνατών προορισμών της κάθε πόλης. | (Tsiotas and Polyzos, 2015c) |
| ($B_3$) Δείκτης ροής υπεραστικού ΚΤΕΛ | Προκύπτει από το γινόμενο των $B_3=B_2 \cdot S_{10}$ | (Tsiotas and Polyzos, 2015c) |
| ($B_4$) Μέση συχνότητα δρομολογίων ΟΣΕ | Ο μέσος όρος του αριθμού των εβδομαδιαίων δρομολογίων ΟΣΕ για το σύνολο των δυνατών προορισμών της κάθε πόλης. | (Tsiotas and Polyzos, 2015c) |
| ($B_5$) Δείκτης ροής δρομολογίων ΟΣΕ | Προκύπτει από το γινόμενο των $B_5=B_4 \cdot S_{11}$. | (Tsiotas and Polyzos, 2015c) |
| ($B_6$) Αριθμός ΙΧ | Αριθμός των ιδιωτικών οχημάτων για κάθε νομό για το έτος 2006. | (Tsiotas and Polyzos, 2015c) |
| ($B_7$) Αριθμός λεωφορείων | Αριθμός των λεωφορείων ΚΤΕΛ ανά νομό για το έτος 2006. | (Tsiotas and Polyzos, 2015c) |
| ($B_8$) Αριθμός ταξί | Αριθμός των ταξί ανά νομό για το έτος 2006. | (Tsiotas and Polyzos, 2015c) |

*Ομάδα Οντολογικών Μεταβλητών (Ontological Class)* - **X**$_O$

| | | |
|---|---|---|
| ($O_1$) Εργατικό πληθυσμιακό δυναμικό | Το ποσοστό του πληθυσμού που αναλογεί στις ηλικίες 20≤Α≤65, με βάση την απογραφή του 2011. | (Tsiotas and Polyzos, 2015c) |
| ($O_2$) Δείκτης εκπαίδευσης | Σύνθετος δείκτης που περιγράφει το επίπεδο εκπαίδευσης του κάθε νομού. | (Πολύζος, 2011) |
| ($O_3$) GDP | Το ακαθάριστο εγχώριο προϊόν (Gross Domestic Product) κάθε νομού για το έτος 2006. | (Πολύζος, 2011; Tsiotas and Polyzos, 2015c) |
| ($O_4$) Δείκτης ευημερίας | Σύνθετος δείκτης που περιγράφει το επίπεδο ευημερίας κάθε νομού για το έτος 2005. | (Tsiotas and Polyzos, 2015c) |
| ($O_5$) Εργατικό δυναμικό δημοσίου | Αριθμός των δημοσίων υπαλλήλων του κάθε νομού για το έτος 2006. | (Tsiotas and Polyzos, 2015c) |
| ($O_6$) Προϊόν του τομέα μεταφορών | Η συμμετοχή κάθε νομού στη διαμόρφωση του προϊόντος στον τομέα των μεταφορών για το έτος 2006. | (Tsiotas and Polyzos, 2015c) |
| ($O_7$) Ατυχήματα | Αριθμός των καταγεγραμμένων από την Τροχαία ατυχημάτων ανά νομό, για το έτος 2006. | (Tsiotas and Polyzos, 2015c) |
| ($O_8$) Κατά κεφαλήν ποσοστό τροχαίων ατυχημάτων | Προκύπτει από το λόγο του αριθμού των ατυχημάτων διά τον πληθυσμό του κάθε νομού ($O_8=O_7/S_6$). | (Tsiotas and Polyzos, 2015c) |
| ($O_9$) Παραγωγικός δυναμισμός | Σύνθετος δείκτης που εξαρτάται από τη μεταβολή του GDP, το ποσοστό ανεργίας, την παραγωγικότητα και το ποσοστό των εργαζομένων. | (Πολύζος, 2011) |
| ($O_{10}$) Ανισότητες στην ανεργία | Τιμές του δείκτη ανισοτήτων του Theil στην ανεργία για το έτος 2007. | (Τσιώτας και Πολύζος, 2012) |

\*. Εντός των παρενθέσεων δίδονται οι συμβολισμοί των μεταβλητών
\*\*. Συμβολίζεται ξεχωριστά επειδή χρησιμοποιείται ως εξαρτημένη μεταβλητή στο υπόδειγμα